\documentclass[twocolumn]{aastex631}

\usepackage{nameref}
\usepackage[normalem]{ulem}
\usepackage{graphicx}
\usepackage{float}

\usepackage{amsmath}

\received{\today}
\newcommand{\angstrom}{$\mathring{\rm A}$}
\submitjournal{\apj}

\graphicspath{{./}{}}

\shorttitle{Variable AGN lags}
\shortauthors{Su et al.}

\begin{document}
\title{Interband Lag Variability in Active Galactic Nuclei across ZTF Data from Multiple Years}

\email{zbsu@mail.ustc.edu.cn; zcai@ustc.edu.cn}

\author[0000-0001-8515-7338]{Zhen-Bo Su}
\affiliation{Department of Astronomy, University of Science and Technology of China, Hefei 230026, People's Republic of China}
\affiliation{School of Astronomy and Space Science, University of Science and Technology of China, Hefei 230026, People's Republic of China}

\author[0000-0002-4223-2198]{Zhen-Yi Cai}
\affiliation{Department of Astronomy, University of Science and Technology of China, Hefei 230026, People's Republic of China}
\affiliation{School of Astronomy and Space Science, University of Science and Technology of China, Hefei 230026, People's Republic of China}

\author[0000-0001-8416-7059]{Hengxiao Guo}
\affiliation{Shanghai Astronomical Observatory, Chinese Academy of Sciences, Shanghai 200030, People's Republic of China}

\author[0000-0002-0771-2153]{Mouyuan Sun}
\affiliation{Department of Astronomy, Xiamen University, Xiamen 361005, People's Republic of China}

\author[0000-0002-4419-6434]{Jun-Xian Wang}
\affiliation{Department of Astronomy, University of Science and Technology of China, Hefei 230026, People's Republic of China}
\affiliation{School of Astronomy and Space Science, University of Science and Technology of China, Hefei 230026, People's Republic of China}
%%%%%%%%%%%%%%%%%%%%%%%%%%%%%%%%%%%%%%%%%%%%%%%%%%%%%%%%%%%%%%%%%%%%%%%%%%%%%%%%

\begin{abstract}
    Interband lags in the optical continua of active galactic nuclei (AGN) have been observed over years of monitoring, yet their physical origins remain unclear.
    While variable interband lags have been found in a few individual AGN potentially, the temporal behavior of interband lags of an AGN sample has not been explored systematically. 
    Here, we analyze the interband lags of 94 bright AGN at $z<0.8$, using both seasonal one-year and full six-year $gri$-band light curves from Zwicky Transient Facility Data Release 22. 
    We find that more than half of 94 AGN show  significant seasonal variations in the interband lags. Besides, the short-term lags, derived by averaging lags inferred from multiple seasonal light curves, are consistently smaller than the long-term lags, which are inferred from the full six-year light curves. 
    This supports recent theoretical simulations where the lag measurement is sensitive to the baseline of light curve and the lag variation could be simply attributed to the inherent randomness of AGN variability. 
    Our findings suggest that the interband lags of AGN are more complex and stochastic than commonly thought, and highlight the importance of high-precision time-domain surveys in uncovering the properties of AGN variability as well as the associated accretion physics.
\end{abstract}

\keywords{accretion, accretion discs - galaxies: active - galaxies: Seyfert - time-domain astronomy}

%%%%%%%%%%%%%%%%%%%%%%%%%%%%%%%%%%%%%%%%%%%%%%%%%%%%%%%%%%%%%%%%%%%%%%%%%%%%%%%%

\section{Introduction} \label{sec:intro}

Active galactic nuclei (AGN) are among the most luminous and enigmatic objects in the universe. They are powered by the accretion of matter onto supermassive black holes (BHs) at their centers, forming an accretion disk that plays a key role in AGN emission. 
According to the standard disk model \citep[SSD, ][]{ShakuraSunyaev1973A}, the disk is optically thick and geometrically thin, emitting radiation predominantly in the ultraviolet (UV) and optical bands \citep{Cai2023NatAs...7.1506C,Cai2024Univ...10..431C}. 
Observations show that UV/optical variability in AGN can span timescales from hours to years \citep{Ulrich1997}. 
These variations often exhibit strong correlations across different wavelengths, with shorter wavelengths typically leading longer ones.
In the context of the accretion disk reprocessing scenario \citep{Cackett2021}, X-ray or far-UV radiation from a central region irradiates the disk, driving variability in the UV/optical bands. 
This process introduces time delays between variations in different bands, which are believed to reflect the light travel time from the central radiation source to the various emission regions of the disk.
As such, interband lags among the optical continua of AGN provide a potential tool for mapping the unresolvable structure of AGN accretion disks, a technique known as continuum reverberation mapping (CRM).

However, the feasibility of inferring disk sizes from interband lags remain inconclusive.
Some studies report that interband lags are 2–10 times larger than theoretical predictions \citep[e.g.,][]{Fau2016,Jiang2017,Fausnaugh2018,Cackett2018,Edelson2019,Jha2022,Guo2022,Montano2022,Kara2023ApJ...947...62K}, while others find that measured lag are consistent with theoretical expectations, albeit with significant scatter \citep[e.g.,][]{Homayouni2019,Yu2020,Sharp2024}. 
These discrepancies suggest that the simple reprocessing scenario alone may be insufficient to account for the UV/optical variability in AGN. 
To reconcile theory with observations, several models have been proposed. Some retain the reprocessing scenario but introduce additional physical process, such as internal reddening within the host galaxy \citep{Gaskell2017MNRAS.467..226G}, deviations from blackbody emission \citep{Hall2018ApJ...854...93H}, contributions from disk winds \citep{Sun2019MNRAS}, the corona-heated accretion-disk reprocessing model \citep{Sun2020ApJ...891..178S}, and diffuse continuum emission (DCE) from the broad line region \citep[BLR,][]{Cackett2018,Chelouche2019,Netzer2022,Montano2022}. 
Alternatively, based on the thermal fluctuation model \citep{DexterAgol2011,Cai2016ApJ...826....7C}, \citet{Cai2018,Cai2020} propose that the interband lags could be the result of the differential regression capability of local thermal fluctuations across the disk.

Additionally, to better understand the origin of interband lags and distinguish between models for AGN variability, \cite{Su2024b} recently simulated multi-band optical light curves for NGC 5548 based on two static variability models, i.e., the thermal fluctuation model \citep{Cai2018,Cai2020} and the reprocessing of both the accretion disk and clouds in the BLR \citep{Jaiswal2023}. Both models predict that the interband lags of an individual AGN would vary randomly with a finite baseline due to the stochastic nature of AGN variability. Notably, the thermal fluctuation model generally predicts greater variation in lags for an individual AGN with repeated, non-overlapping observations compared to the reprocessing model. This highlights that variation in interband lags could help understand their origin.

In observations, repeated monitoring campaigns have revealed potential varied lag in several individual AGN, such as NGC 7469 \citep{Vincentelli2023}, NGC 4395 \citep{Montano2022, McHardy2023MNRAS.519.3366M}, Fairall 9 \citep{Pal2017MNRAS.466.1777P, Santisteban2020, Edelson2024} and NGC 4151 \citep{Zhou2024arXiv}.
Additionally, using multi-season light curves from the Dark Energy Survey (DES), \cite{Yu2020} reported that one quasar in a sample of seven exhibited a varying disk size, whereas \cite{Mudd2018} found that the consistency of disk sizes across seasons remained uncertain, with variations generally within 1$\sigma$–1.5$\sigma$ for five quasars. 
However, up to now, only a few local AGN have been monitored yearly by campaigns, such as the AGN Space Telescope and Optical Reverberation Mapping project \citep[STORM;][]{DeRosa2015ApJ...806..128D}. 
In addition, the limited number of observing seasons ($\sim$2–3) and the small AGN sample in DES observations \citep{Mudd2018,Yu2020} make it difficult to draw firm statistical conclusions about whether lags exhibit randomness. 
Therefore, a systematic investigation into the randomness of interband lags is highly warranted.

The Zwicky Transient Facility (ZTF) surveys have provided high-cadence ($\sim$1-3 day) and long-term photometric data, offering a unique opportunity to analyze the interband lags both within individual observing seasons (i.e., $\sim$1 year) and across the full span of several years covered by a given ZTF data release. 
Previous studies using the ZTF data have typically focused on either single observing season or the entire available baseline, consistently reporting lags larger than those predicted by the SSD model.
For example, \cite{Jha2022} cross-matched reverberation-mapped AGN with the ZTF sixth data release, selecting a sample of 19 AGN observed in two seasons but focusing only on the season with the most reliable lag measurements. 
Similarly, \cite{Guo2022} analyzed a sample of 92 AGN at $z < 0.75$, limiting their analysis to the first-season light curves from the ZTF third data release. 
More recently, \cite{GuoH2022} investigated 94 AGN at $z < 0.8$, utilizing the whole light curves spanning $\sim4$ years using the ZTF Seventh Data Release (DR7).
Despite the availability of multi-year ZTF observations, the variability of lags across different observing seasons remains unexplored.

In this work, benefiting from the $\sim$six-year photometric data provided by the ZTF Twenty-second Data Release (DR22), we conduct a systematic investigation into potential seasonal variations in interband lags for a sample of 94 AGN with significant lag detections reported by \cite{GuoH2022}.
We introduce the data and methodology in Section~\ref{sec:sample}, present statistical results and interpret the implications of the measured interband lags in Section~\ref{sec:results}, and conclude with a brief summary in Section~\ref{sec:cons}.

\section{Data and Methods}\label{sec:sample}

\subsection{AGN sample and ZTF light curves}

\begin{figure*}[ht]
    \centering
    \includegraphics[width=0.6\textwidth]{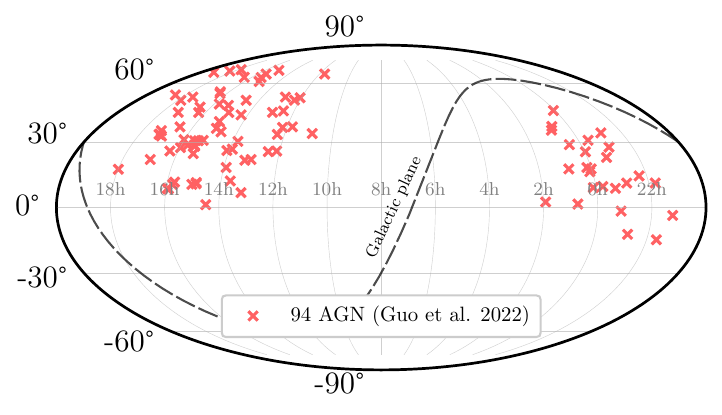}
    \includegraphics[width=0.33\textwidth]{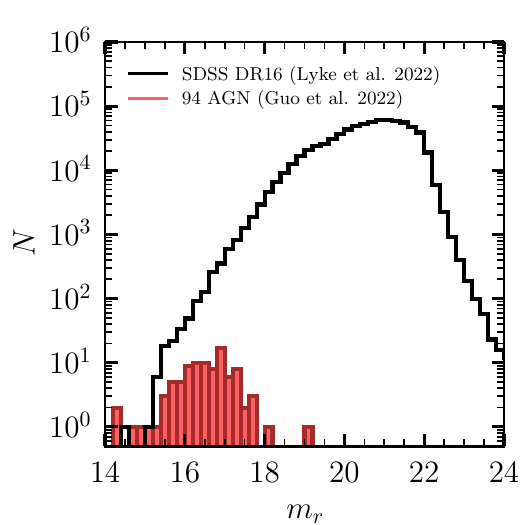}
    \caption{Left panel: distribution on the sky in equatorial coordinates of the 94 AGN with significant interband lags reported by \cite{GuoH2022}.
    Right panel: distributions of the apparent $r$-band magnitude for the 94 AGN (\citealt{GuoH2022}, red filled histogram) and $\simeq 0.75$ million SDSS DR16 quasars \citep[][black histogram]{SDSSdr16q}.}
    \label{fig:coord_mag}
\end{figure*}

\begin{figure*}
    \centering
    \includegraphics[width=0.24\textwidth]{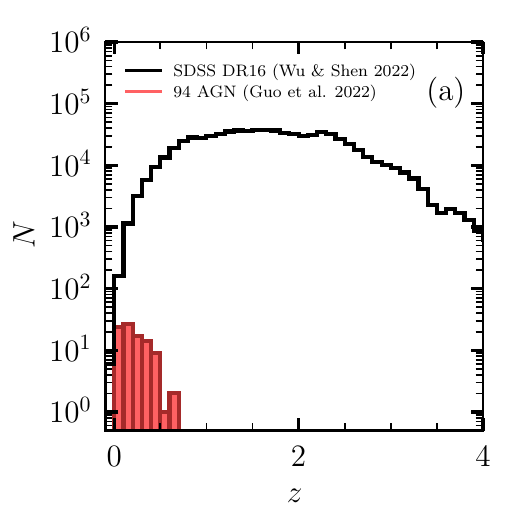}
    \includegraphics[width=0.24\textwidth]{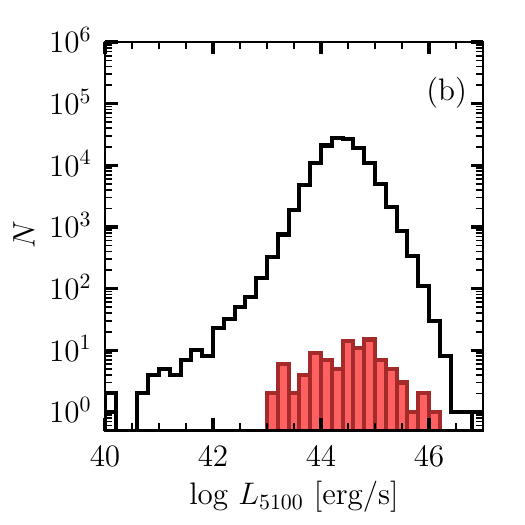}
    \includegraphics[width=0.24\textwidth]{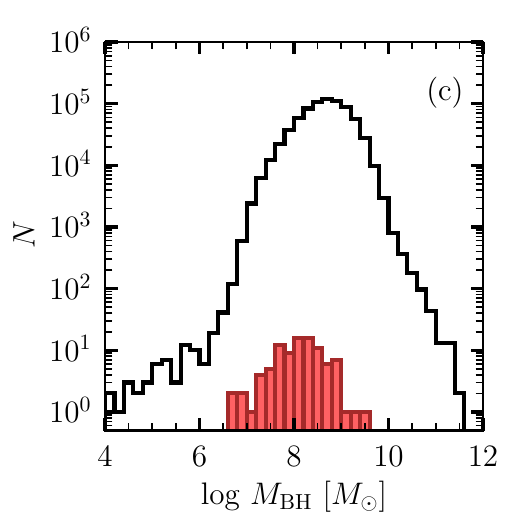}
    \includegraphics[width=0.24\textwidth]{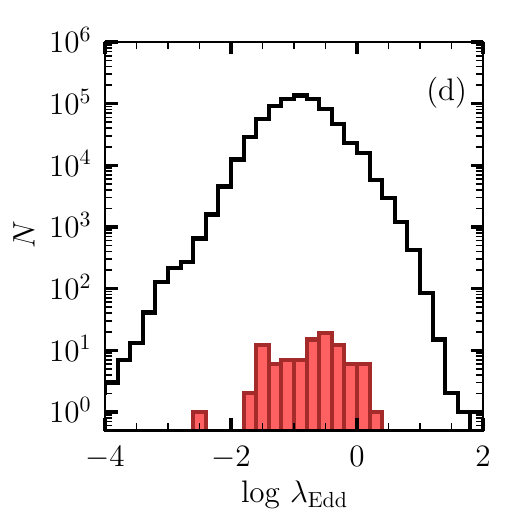}
    \caption{From left to right panels, distributions of four physical properties, i.e., redshift ($z$),  5100~\angstrom~monochromatic luminosity ($L_{5100}$), BH mass ($M_{\rm BH}$), and Eddington ratio ($\lambda_{\rm Edd}$), for the 94 AGN (red filled histogram) and the whole SDSS DR16 quasars (black histogram).
    The typical (i.e., median and 16-84th percentile ranges) properties of the 94 AGN are $z = 0.18^{+0.17}_{-0.09}$,  $\log L_{5100} = 44.56^{+0.59}_{-0.72}$, $\log M_{\rm BH} = 8.2^{+0.43}_{-0.56}$, and $\log \lambda_{\rm Edd} = -0.63^{+0.38}_{-0.75}$.
    }
    \label{fig:agn_prop}
\end{figure*}

In this work, we begin with a sample of 94 AGN at $z < 0.8$ from \cite{GuoH2022}, for which high-quality and significant interband lags were measured using the ZTF DR7 light curves.  
These AGN are selected based on the quality assessment of the lag measurements, consistency between lag measurements obtained from two independent methods, and lag uncertainties, among others (cf. Table 1 and Table 2 in \citealt{GuoH2022}).
To explore the variation in lags across different observing seasons, we extend the ZTF DR7 light curves by incorporating an additional 2 years of light curves from the ZTF DR22\footnote{\url{https://irsa.ipac.caltech.edu/data/ZTF/docs/releases/dr22/ztf_release_notes_dr22.pdf}}. 
To exclude effects of bad pixels, clouds, and the Moon on the photometric data, we apply \texttt{catflags=0} to remove those unusable observation epochs.

The left panel of Figure~\ref{fig:coord_mag} shows the distribution of the 94 AGN on the sky map. As shown in the right panel of Figure~\ref{fig:coord_mag}, due to the shallow $r$-band magnitude detection limit of ZTF, with a 5$\sigma$ detection limit of 20.6 mag \citep{Masci2019}, only the brighter AGN, with apparent $r$-band magnitudes $m_r \gtrsim 18$, have reliable lags measured in \cite{GuoH2022}. 
In contrast, many fainter AGN, e.g., in the whole SDSS DR16 quasars, have been studied less extensively.

Figure~\ref{fig:agn_prop} illustrates distributions of physical properties for these 94 AGN and the whole SDSS DR16 quasars. 
It is evident that the interband lags of a significant number of AGN remain unexplored.
The ongoing wide-field time-domain surveys, such as the 2.5-meter Wide Field Survey Telescope \citep[WFST,][]{wfst2023}, are expected to play a crucial role in uncovering interband lags for a large population of fainter AGN and in advancing our understanding of the relationship between interband lags and physical properties.

\subsection{Visibility and observing seasons}

\begin{figure*}
    \centering
    \includegraphics[width=0.95\textwidth]{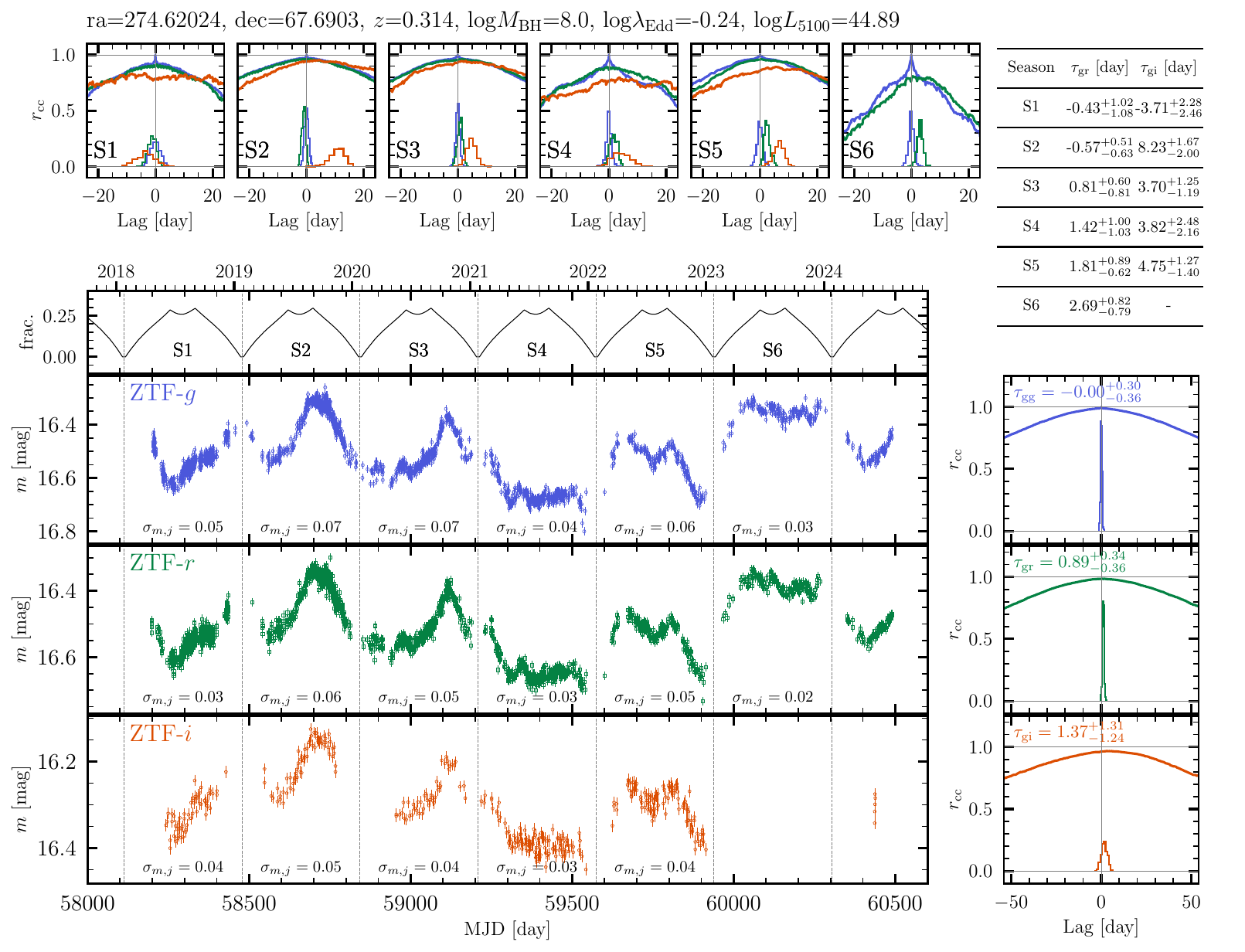}
    \caption{Illustration of the ZTF $gri$-band light curves (main panel) and the interband CCF results (panels in the top row and the right column) for a typical AGN in the sample.
    In the main panel, according to the minimal visibility of the AGN over years (vertical dashed lines), the ZTF $g$- (blue circles), $r$- (green triangles), and $i$-band (orange pentagons) light curves are subdivided into six seasons. 
    In the $j$-th season, the seasonal variation amplitude of magnitude, $\sigma^{\rm K24}_{m,j}$ (Equation~(\ref{eq:var1})), is nominated.
    For each season, the top-row panel presents the seasonal CCF results (curves) and lag distributions (histograms) for the $g$ (blue), $r$ (green), and $i$ (orange) bands relative to the $g$ band, while the lags in the restframe and their 1$\sigma$ uncertainties, estimated using FR method, are tabulated in the top-right panel. 
    Note that the lag of the $g$ band relative to itself, $\tau_{\rm gg}$, is identically zero and not shown.
    The right-column panel shows the corresponding CCF results (curves) and lag distributions (histograms) for the full six-year light curves, while the lag in the restframe and their 1$\sigma$ uncertainties, estimated using the FR method, are labeled in each plot.
    Note that we search the centroid lag in the observed frame within $\pm 50$ days for the one-year light curves (the top-row panels) and $\pm 300$ days for the six-year light curves (the right-column panels), wider than the abscissa ranges shown.}
    \label{fig:ztf_lc_ccf}
\end{figure*}

\begin{figure*}
\centering
\includegraphics[width=0.3\linewidth]{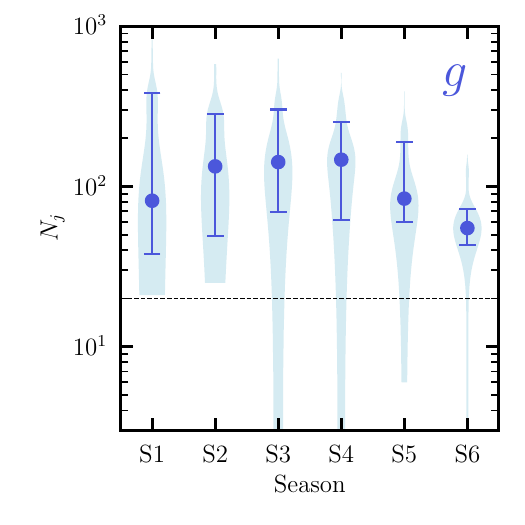}
\includegraphics[width=0.3\linewidth]{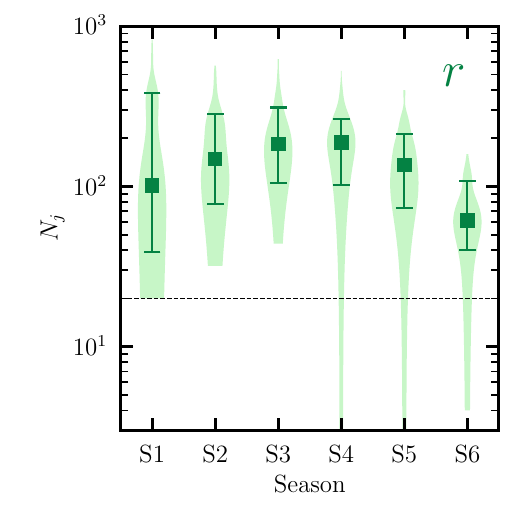}
\includegraphics[width=0.3\linewidth]{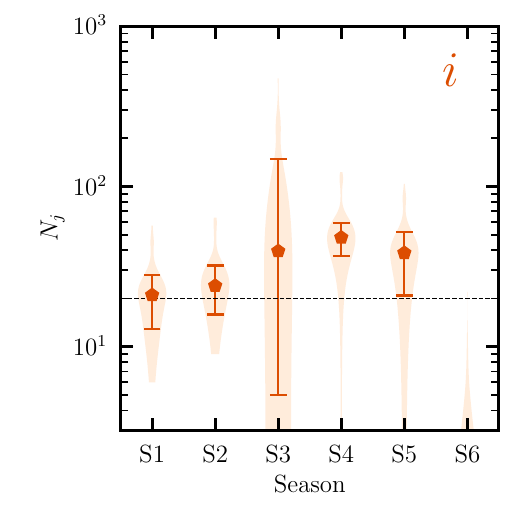}
\includegraphics[width=0.3\linewidth]{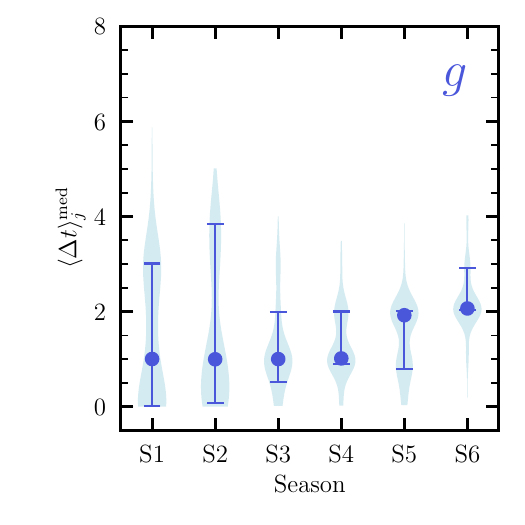}
\includegraphics[width=0.3\linewidth]{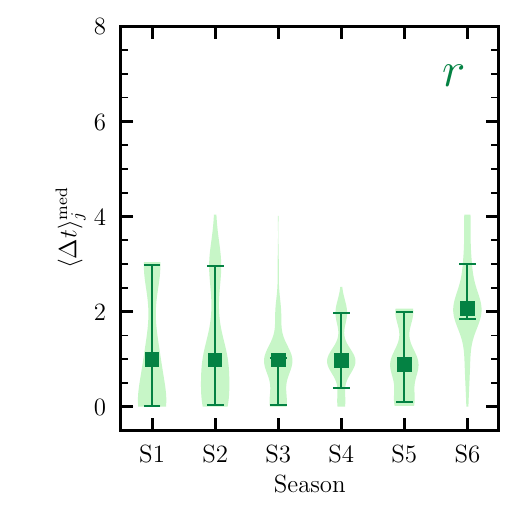}
\includegraphics[width=0.3\linewidth]{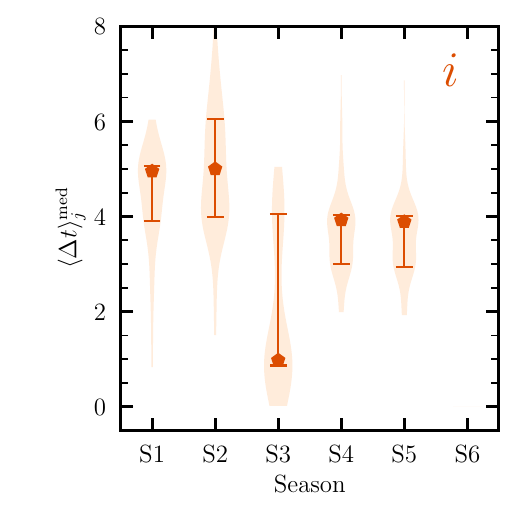}
\includegraphics[width=0.3\linewidth]{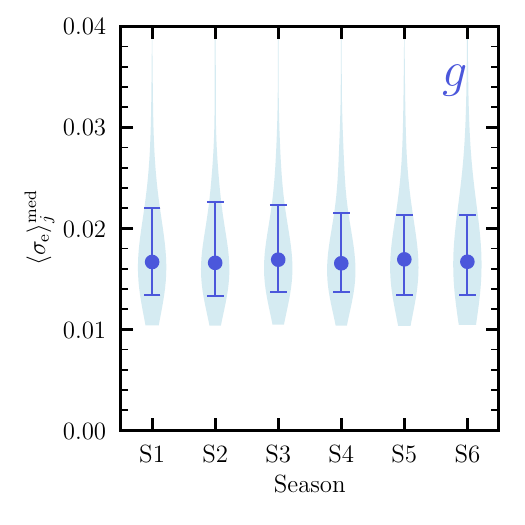}
\includegraphics[width=0.3\linewidth]{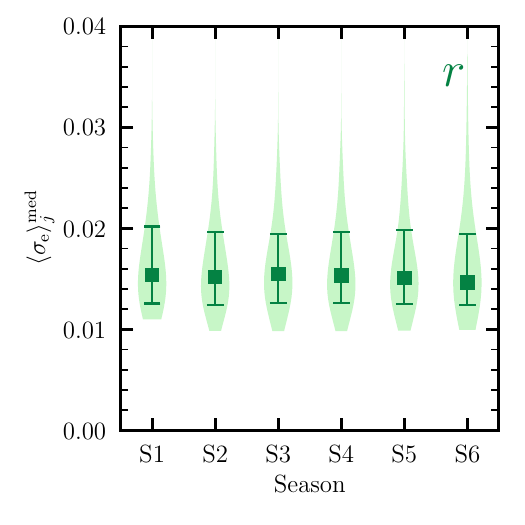}
\includegraphics[width=0.3\linewidth]{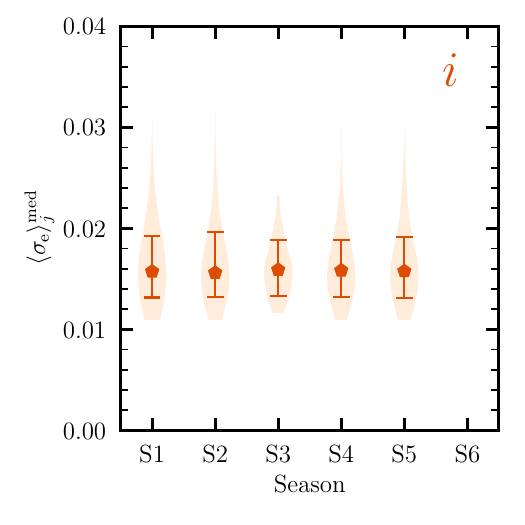}
\caption{
Three observational characteristics, i.e., the number of observed epochs ($N_j$; top panels), the median cadence ($\langle \Delta t\rangle^{\rm med}_j$; middle panels), and the median photometric uncertainty ($\langle \sigma_{\rm e}\rangle^{\rm med}_j$; bottom panels) of each AGN observed in the $j$-th season and in the ZTF $g$ (left column), $r$ (middle column), and $i$ (right column) bands. 
In each plot and each season, the violin plot illustrates the distribution for the 94 AGN, while the symbol indicates the median superimposed by the 16\%$-$84\% percentile range of the distribution.
}
\label{fig:agn_season_obsinfo}
\end{figure*}

\begin{figure*}[ht]
\centering
\includegraphics[width=.95\linewidth]{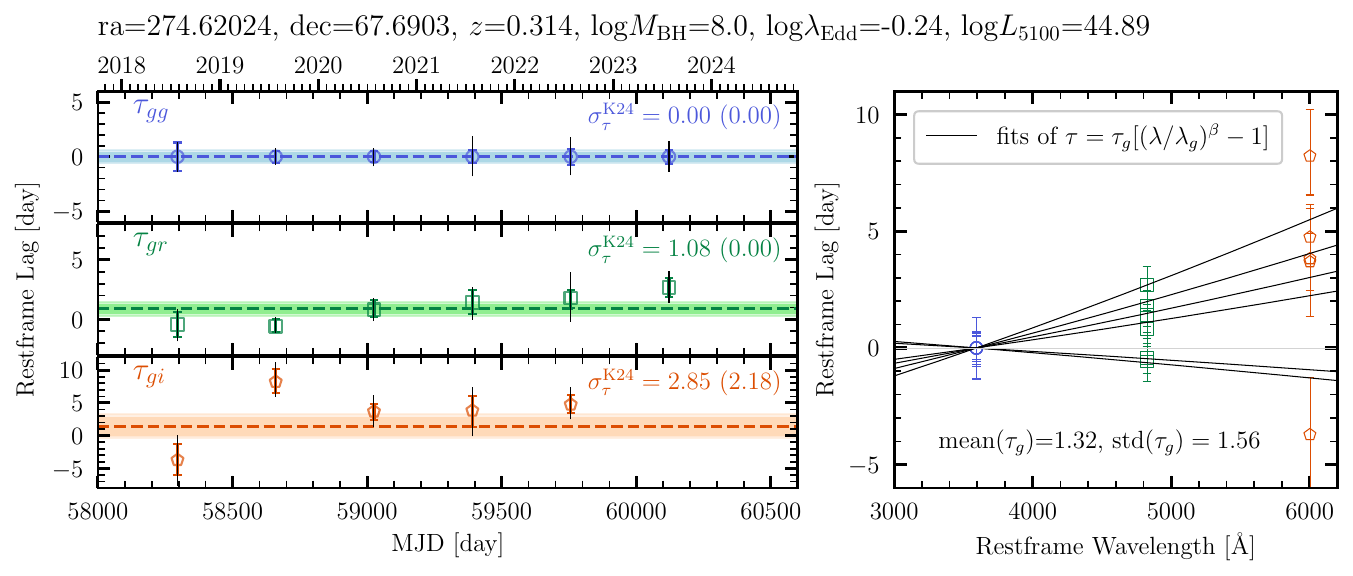}
\caption{
    Illustration of the rest-frame interband lags measured in six seasons (left panel) and the annual lag-wavelength relations (right panel) for the same AGN shown in Figure~\ref{fig:ztf_lc_ccf}. 
    Left panel: the seasonal centroid lags, $\tau_{gx}$, of an $x$ band relative to the $g$ band, where $x$ is $g$, $r$, or $i$ band. Using the 1$\sigma$ FR (thick error bars) and FR/RSS (thin error bars) uncertainties of the centroid lag, we report the scatter in the seasonal lag as $\sigma^{\rm K24}_\tau$ (Equation~(\ref{eq:var1})).
    The smaller value of $\sigma^{\rm K24}_\tau$ in brackets results from the larger 1$\sigma$ FR/RSS uncertainties.
    For comparison, utilizing the full six-year light curves, the horizontal lines and shaded areas are the centroid lags and their 1$\sigma$ uncertainties, respectively. 
    Right panel: by fitting a functional form of $\tau = \tau_g[(\lambda/\lambda_g)^{\beta}-1]$ with $\beta = 4/3$ to each lag-wavelength relation observed, the best-fit $\tau_g$ is commonly used to infer the $g$-band absolute disk size. However, the lag-wavelength relations as well as the best-fit $\tau_g$, whose mean and standard deviation (std) are nominated, are clearly variable across six seasons.}
\label{fig:agn_lag}
\end{figure*}

Interband lags in AGN are critical to the physical models of accretion disks, and obtaining meaningful constraints on the accretion physics require first establishing whether the lags remain stable over time.
If the lags are indeed stable, it would support the use of lags measured within a limited baseline to infer disk properties such as the disk size. 
If not, it raises important questions about the physical mechanisms responsible for the observed variations.
To date a few AGN have shown potential variation in lags (see~\nameref{sec:intro}), but most of them lack multi-year repeated monitoring. 
Currently, the ZTF surveys offer
high-cadence ($\sim 1-3$ days sampling) and long-term ($\sim$ 6 years) photometric data, providing a unique opportunity to investigate the lag variability across different observing seasons. Here we define an observing season as a period that has a length of one year.

To investigate whether interband lags vary over time, we divide the light curve of each individual AGN into six seasonal segments, each corresponding to $\simeq1$ year of observations. 
The segmentation is guided by the seasonal gaps in the light curve, which arise from the orbital motion of the Earth.
As the Earth moves around the Sun, the observable night sky changes throughout the year, thus an astronomical object is observable only during a specific period. 
To perform the segmentation, we use the \texttt{Astroplan} \citep{astroplan} package, which is designed to assist in scheduling astronomical observations.
By providing the geodetic coordinates (longitude, latitude, and altitude) of the Palomar Observatory, we apply constraints such as airmass ($\leqslant 2$), altitude ($\geqslant 30^{\circ}$), and astronomical twilight to replicate the real-world observation conditions. This approach allow us to determine the visibility of each AGN observable every year from the Palomar Observatory. 

The top row of the main panel of Figure~\ref{fig:ztf_lc_ccf} illustrates the fractional visibility as a function of time for an AGN located at RA=274.62024$^\circ$ and DEC=67.6903$^\circ$. 
The AGN, whose $z$=0.314, $\log M_{\rm BH}$=8.0, $\log \lambda_{\rm Edd}$=-0.24, and $\log L_{5100}$=44.89, is a typical AGN within the sample of 94 AGN.
The seasonal gap is identified by the minimal visibility over the years, as indicated by the vertical dotted lines in the same panel.
For the ZTF-DR22, the $g$- and $r$-band light curves span from March 2018 to July 2024, covering $\simeq 6.5$ observing seasons. In contrast, the $i$-band light curve only spans from March 2018 to March 2023, covering $\simeq5$ observing seasons.
Therefore, we focus only on the first six observing seasons in the DR22.
For each season, such as the $j$-th season ($j = 1 \ldots 6$), we present several observational characteristics, including the number of observed epochs, $N_j$, the median cadence, $\langle \Delta t \rangle^{\rm med}_j$, and the median photometric uncertainties, $\langle \sigma_{\rm e} \rangle^{\rm med}_j$ for the $k$-th AGN ($k = 1 \ldots 94$) observed annually in the ZTF $x$ band, where $x \in \{g,~r,~i\}$.

Figure~\ref{fig:agn_season_obsinfo} presents the seasonal observation information for the 94 AGN.
In the top panel, we find that in each season most AGN are observed as many as $\sim 100$ epochs in the $g$ and $r$ bands, while in the $i$ band some of them are observed with less than 20 epochs, especially in the first two seasons and the last one.
In the middle panel, the median value of cadence $\langle \Delta t \rangle^{\rm med}_j$ is $\simeq 1$ day for the $g$ and $r$ bands, while a lower median cadence in the $i$ band.
In the bottom panel, the photometric uncertainties for the ZTF-$gri$ bands are stable for the 94 AGN, with median values of $\langle \sigma_{\rm e}\rangle^{\rm med}_j$ of $\simeq$ 0.016 mag, $\simeq$ 0.015 mag, and $\simeq$ 0.016 mag for the $g$, $r$, and $i$ bands, respectively.

\subsection{Interband lags inferred from seasonal one-year and full six-year light curves}\label{sec:lags}

\begin{figure*}[ht]
    \centering
    \includegraphics[width=0.4\textwidth]{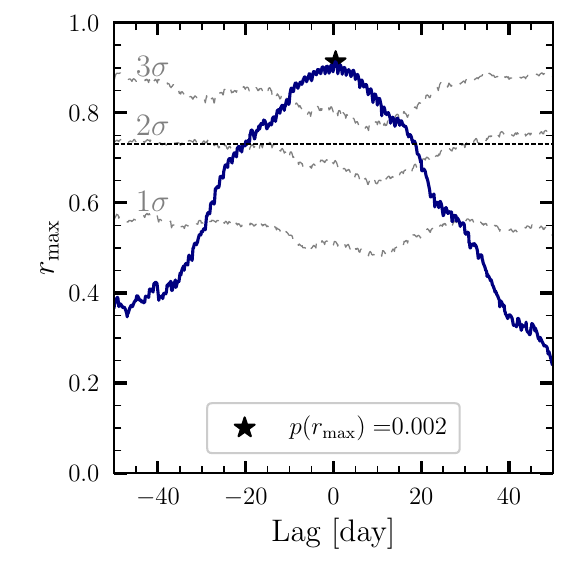}
    \includegraphics[width=0.4\textwidth]{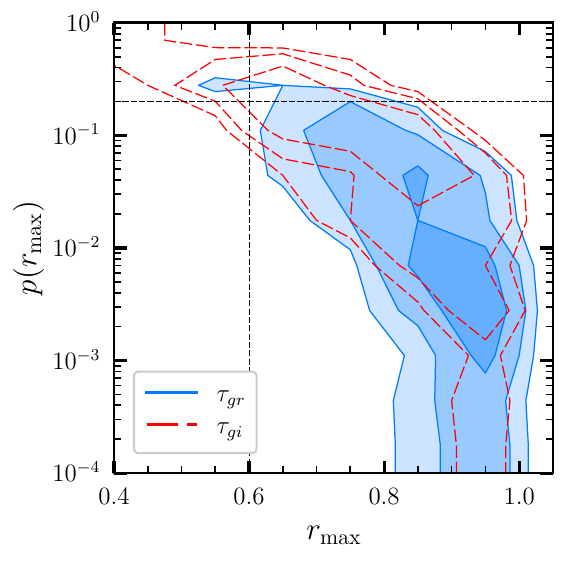}
    \caption{Left panel: an example of the observed CCF between the first seasonal $g$- and $r$-band light curves of the same AGN shown in Figure~\ref{fig:ztf_lc_ccf}. The three gray dashed curves from bottom to top are the $1\sigma$ to $3\sigma$ levels of the CCF, respectively. $p(r_{\rm max})$ is the significance of $r_{\rm max}$.
    Right panel: contour plots of $p(r_{\rm max})$ versus $r_{\rm max}$ for $\tau_{gr}$ (blue solid contours) and $\tau_{gi}$ (red dashed contours) of the 94 AGN. 
    The bottom-right region outlined by the two dashed lines, i.e., $r_{\rm max}>0.6$ (vertical) and $p(r_{\rm max})<0.2$ (horizontal), represents our definition of a success lag, following \citet{U2022}.
    }
    \label{fig:rmax_pvalue}
\end{figure*}

We measure the interband lags using both seasonal one-year and full six-year light curves.
The interband lags of the $g$, $r$, and $i$ bands relative to the $g$ band are computed using the standard interpolated cross-correlation function (CCF) method. 
For the seasonal one-year light curves, we search for lags within a range of -50 to 50 days to ensure sufficient coverage and adopt the traditional centroid lag, defined as the CCF-weighted average lag with correlation coefficients, $r_{\rm cc}$, exceeding 80\% of its maximum value, $r_{\rm max}$, i.e., $r_{\rm cc} \geqslant 0.8~r_{\rm max}$. 
For the full six-year light curves, the CCF become very broad, with a flat top spanning several hundred days.
Therefore, the lag search range is expanded to -300 to 300 days, within which the CCF with $r_{\rm cc} \geqslant 0.8~r_{\rm max}$ is fully covered. 
Throughout our analysis, we adopt the centroid lag weighted with $r_{\rm cc} \geqslant 0.8~r_{\rm max}$, as it could represent the luminosity-weighted radius of the emission region 
if assuming the reprocessing scenario \citep[][see also \citealt{Goad2014}]{Koratkar1991}. But, we complement our results with the centroid lag weighted with $r_{\rm cc} \geqslant 0.95~r_{\rm max}$ following \cite{GuoH2022}. 
To estimate the uncertainty of the lag, we use both the flux randomization (FR) and the flux randomization/random subset selection (FR/RSS) methods \citep{Fau2016}. A distribution of the centroid lags is generated from 10$^3$ realizations, and the $1\sigma$ dispersion of this distribution is adopted as the uncertainty of the lag. 

In Figure~\ref{fig:ztf_lc_ccf}, the top (right) panel shows the CCF and lag distribution measured using the seasonal one-year (full six-year) light curves of the specific AGN, while the corresponding centroid lags and $1\sigma$ FR (FR/RSS) uncertainties are depicted in Figure~\ref{fig:agn_lag}, where the FR/RSS uncertainties are generally larger than the FR ones. 
Note that we would only consider the lags inferred from more than 20 epochs per season per band in the following to alleviate the effect of uncertain lags inferred from light curves with too few epochs.
The selection of the minimal epoch number is guided by the fact that the minimal median epoch number is $\sim 20$ for the 94 AGN observed in the $i$ band over seasons (Figure~\ref{fig:agn_season_obsinfo}).

Next, we assess the reliability of the lag measurement following \citet[][\texttt{PyI$^2$CCF} \footnote{https://github.com/legolason/PyIICCF/}]{I2CCF} and \citet{U2022}. 
For each AGN, we model its ZTF $g-$, $r-$, and $i-$band full six-year light curves by a damped random walk (DRW) process using \texttt{celerite} \citep{celerite} and infer the underlying DRW parameters, i.e., damping timescale and variation amplitude.
Using the DRW parameters of each band, we simulate $5000$ ideal (i.e., 0.1 day cadence and 6 year baseline) light curves, which are resampled by linear interpolation such that they share the same observed characteristics, including duration, cadences, and photometric uncertainties. For any two bands, e.g., $g$ and $r$, we obtain $5000$ simulated CCFs of the observed $g$-band light curve relative to $5000$ simulated $r$-band light curves, and another $5000$ simulated CCFs of the observed $r$-band light curve to those simulated in the $g$ band. 
Based on these $2 \times 5000$ simulated CCFs, we derive the $1\sigma$, $2\sigma$, and $3\sigma$ confidence levels for $r_{\rm cc}$ as a function of the lag.
To assess the significance of the observed maximum correlation coefficient, $r_{\rm max}^{\rm obs}$, we calculate the probability of a simulated CCF having a maximum correlation coefficient, $r_{\rm max}^{\rm sim}$, exceeding $r_{\rm max}^{\rm obs}$ as
\begin{equation}
    p(r_{\rm max}) = N(r_{\rm max}^{\rm sim} > r_{\rm max}^{\rm obs})/N(r_{\rm max}^{\rm sim}),
\end{equation}
where $N(r_{\rm max}^{\rm sim}) = 10^4$ and $N(r_{\rm max}^{\rm sim} > r_{\rm max}^{\rm obs})$ is the total number of simulated CCFs with $r_{\rm max}^{\rm sim} > r_{\rm max}^{\rm obs}$. This quantifies the likelihood of the observed correlation arising by chance.
As an example, the left panel of Figure~\ref{fig:rmax_pvalue} shows the observed CCF between the $g$- and $r$-band light curves in the first season for the AGN displayed in Figure~\ref{fig:ztf_lc_ccf}.
The $1\sigma$, $2\sigma$, and $3\sigma$ confidence levels derived from the simulated CCF distribution are represented by gray dashed curves. 
The peak value of the observed CCF, $r_{\rm max}$, highlighted by an asterisk, has a probability $p(r_{\rm max}) = 0.2\%$.
The above simulation is performed for any two seasonal one-year (as well as full six-year) light curves of all 94 AGN.

The reliability of all seasonal lags for the 94 AGN is presented as contour plots of $p(r_{\rm max})$ versus $r_{\rm max}$ in the right panel of Figure~\ref{fig:rmax_pvalue}. 
A clear anti-correlation between $r_{\rm max}$ and $p(r_{\rm max})$ is observed. 
Following \cite{U2022}, we define a successful lag by requiring $r_{\rm max}>0.6$ and $p(r_{\rm max})<0.2$ for all AGN, but we caution that the proper thresholds of $r_{\rm max}$ and $p(r_{\rm max})$ may differ from one AGN to another.

\begin{figure*}[ht]
\centering
\includegraphics[width=0.3\linewidth]{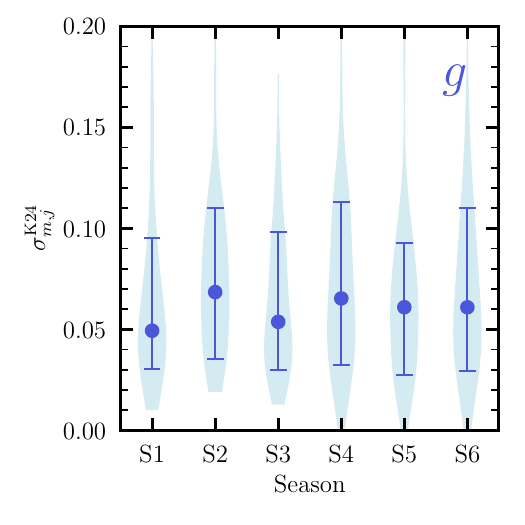}
\includegraphics[width=0.3\linewidth]{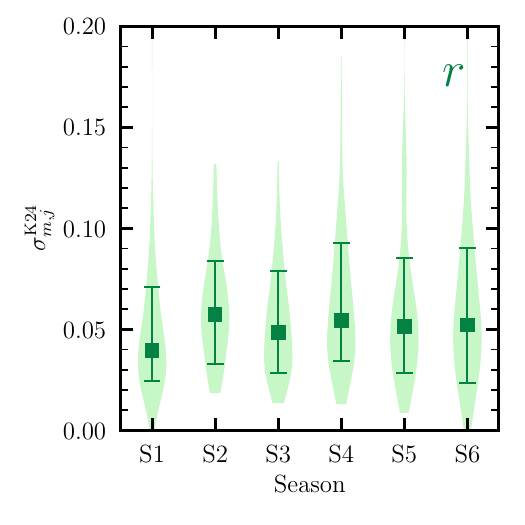}
\includegraphics[width=0.3\linewidth]{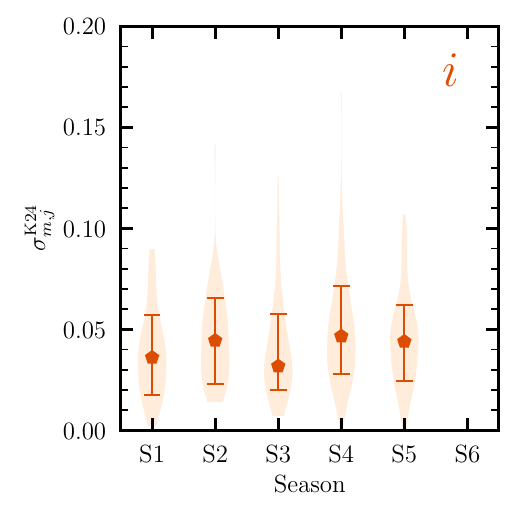}
\includegraphics[width=0.3\linewidth]{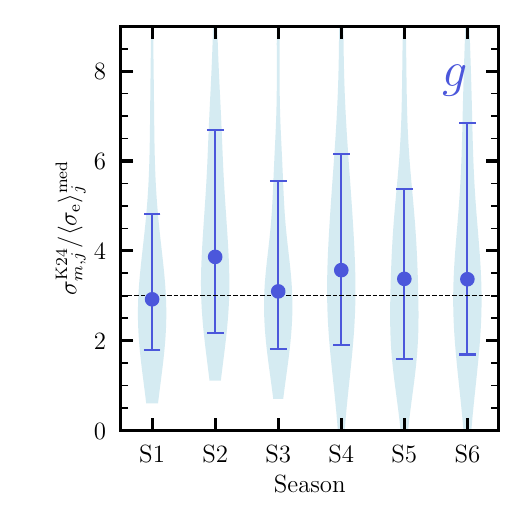} 
\includegraphics[width=0.3\linewidth]{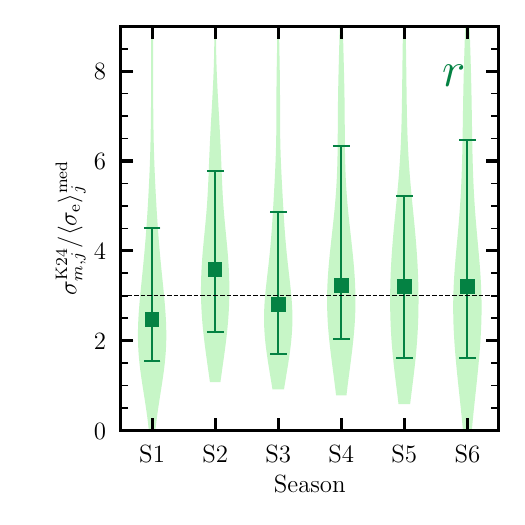}
\includegraphics[width=0.3\linewidth]{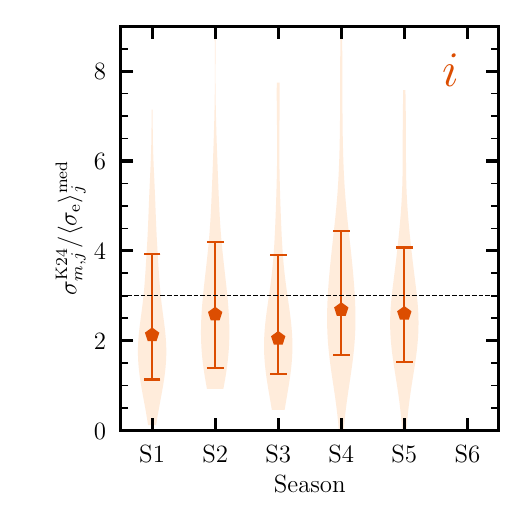}
\caption{
Same as Figure~\ref{fig:agn_season_obsinfo}, but for the optical variation properties, i.e., the variation amplitude ($\sigma^{\rm K24}_{m,j}$; top panels) and the variation significance ($\sigma^{\rm K24}_{m,j}/\langle \sigma_{\rm e} \rangle^{\rm med}_j$; bottom panels). 
In each bottom panel, the horizontal dashed line indicates the $3\sigma$ level.}
\label{fig:agn_var_snr}
\end{figure*}

\begin{figure*}[ht]
\centering
\includegraphics[width=0.3\linewidth]{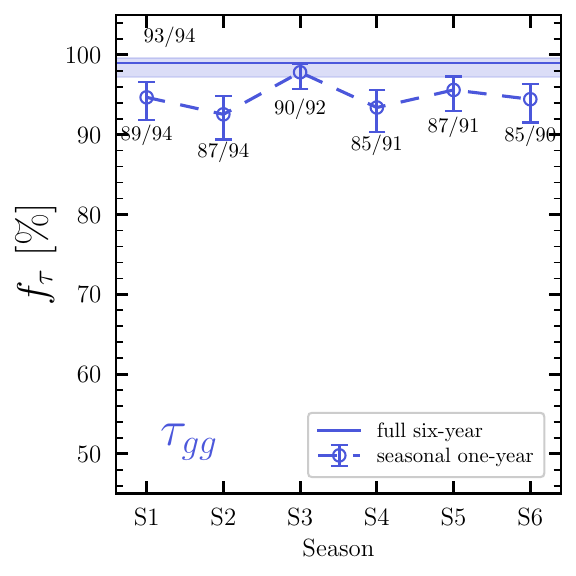}
\includegraphics[width=0.3\linewidth]{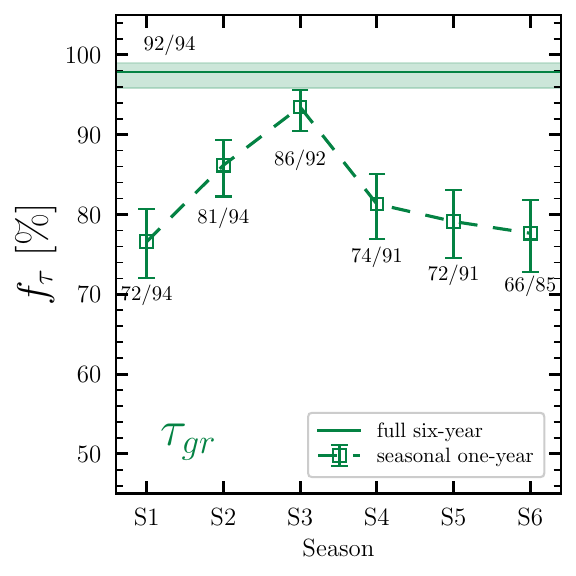}
\includegraphics[width=0.3\linewidth]{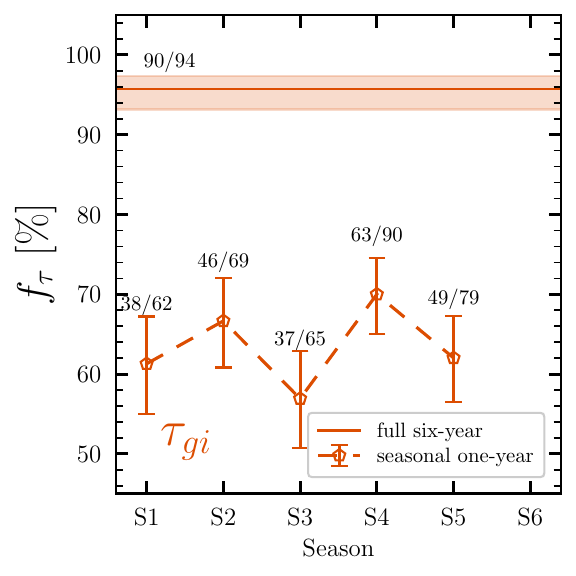}
\caption{
Illustration of the success rate of the lag measurement for the $g$ (left panel), $r$ (middle panel), and $i$ (right panel) bands relative to the $g$ band for sub-samples of the 94 AGN selected with more than observed 20 epochs per season in both bands. The success rates inferred from the seasonal one-year light curves are shown as symbols connected by dashed lines, with vertical error bars indicating the 1$\sigma$ confidence intervals estimated using the Wilson score method \citep{wilson1927}. The horizontal solid line marks the success rate based on the full six-year light curve, and the surrounding shaded region denotes its 1$\sigma$ confidence interval, also estimated using the Wilson score method. 
Near each symbol or the horizontal line, numbers of the corresponding sub-sample (denominator) and of AGN with a successful lag (numerator) are marked.
}
\label{fig:agn_frac}
\end{figure*}

\subsection{Variation amplitude of Magnitude and Lag}\label{sec:amplitude_lag}

To quantify the variation amplitude in both the magnitude ($\sigma_m$) and the interband lag ($\sigma_\tau$), we use the weighted excess variance following \citet[][]{Kang2024}. 
The weighted excess variance is given by
\begin{equation}
     \sigma^{\rm K24}_{m/\tau} = \sqrt{\frac{1}{\sum_j \sigma_j^w} \sum_j \sigma_j^w \left[ \frac{N}{N-1}(X_j - \bar{X})^2 - \sigma_j^2 \right]},
     \label{eq:var1}
\end{equation}
where $X_j$ represents either the magnitude $m_j$ or the lag $\tau_j$, and $\sigma_j$ is the associated uncertainty, in the $j$-th season. 
$\bar{X}$ is the weighted mean value of $X_j$, where the weighted factor is represented by $\sigma_j^w$ with $w = -2$ throughout this work. If $w = 0$, it returns to the canonical excess variance proposed by \citet{Vaughan2003MNRAS}. Using a more negative $w$ means decreasing the contribution of the lags with large uncertainties to the lag variation and would result in a smaller variation amplitude for lags.
To some extent the $\sigma^{\rm K24}_{m/\tau}$ estimator takes the measurement uncertainties into account. If the measurement uncertainties are so large that a negative result is found under the square root of Equation~(\ref{eq:var1}), we set $\sigma^{\rm K24}_{m/\tau}=0$. 
In Section~\ref{sec:var_lag}, we would incorporate two additional methods to quantify the lag variation.
For the AGN shown in Figure~\ref{fig:ztf_lc_ccf}, the variation amplitude of the magnitude in each season is nominated in the main panel of Figure~\ref{fig:ztf_lc_ccf}, while the scatter of the lags among six seasons is nominated in the left panel of Figure~\ref{fig:agn_lag}, showcasing a variable behavior. 

To quantify the lag-wavelength relation measured in each season, we adopt the widely used functional form, i.e., $\tau = \tau_g \left[ \left( \lambda/\lambda_g \right)^{\beta} - 1 \right]$ with a fixed parameter $\beta = 4/3$ predicted by the SSD model. 
According to the continuum reverberation mapping, which assumes the X-ray or far-UV radiation originates from the BH vicinity and drives the variability in longer UV/optical wavelengths, the absolute size of the disk in $g$ band can be inferred as $R_g=c\tau_g$ \citep{Fau2016}, where $c$ is the speed of light and $\tau_g$ is the best-fit parameter from the lag-wavelength relation.
Here we use the best-fit parameter $\tau_g$ as a proxy to quantify the trend of lag-wavelength relations measured in different seasons. 
{However, it should be cautious that the validity of using the interband lag to estimate the disk size has not yet been justified} (see discussion in Section~\ref{sec:var_lag}).

The right panel of Figure~\ref{fig:agn_lag} shows the rest-frame lag-wavelength relations, where the best fits for individual seasons are plotted as black solid curves. 
We find that the lag-wavelength relations are diverse rather than stable. Additionally, we in Figure~\ref{fig:agn_lag} observe two negative $\tau_g$ out of six for the AGN shown in Figure~\ref{fig:ztf_lc_ccf}, albeit with large measurement uncertainties. 
The negative lag is expected by the thermal fluctuation model for AGN variability \citep{Cai2018,Su2024b}.

\subsection{Variation Amplitude, Variation Significance, and the Success Rate of Lag Measurement}

Using Equation~(\ref{eq:var1}), we calculate the $gri$-band variation amplitudes of the magnitude, $\sigma_{m,j}^{\rm K24}$, for the 94 AGN in the $j$-th season.
The results are shown in the top panel of Figure~\ref{fig:agn_var_snr}.
In the bottom panel of Figure~\ref{fig:agn_var_snr}, we present the variation significance, defined as $\sigma_{m,j}^{\rm K24}/\langle \sigma_{\rm e} \rangle_j$, for each AGN observed in the $j$-th season. 
We find that most $g$- and $r$-band light curves exhibit significant variability, with $\sigma_{m,j}^{\rm K24}/\langle \sigma_{\rm e} \rangle_j \gtrsim 3$. 
In contrast, the $i$-band light curves show less significant variability.

Based on the definition of a successful lag in Section~\ref{sec:lags}, we examine the success rate of lag measurements derived from both the seasonal one-year and full six-year light curves of the 94 AGN in Figure~\ref{fig:agn_frac}. 
Note that for any lag between two bands and in each season, we only consider a sub-sample of the 94 AGN with more than 20 epochs per season in both bands to determine the success rate of lag.
As a result of a larger number of observed epochs and a smaller median cadence of $\sim 1$ day, AGN observed in the $g$ and $r$ bands achieve a higher fraction of successful lag measurements. 
In contrast, AGN observed in the $i$ band show a lower success rate owing to a fewer number of observe epochs and a longer median cadence of $\sim 4$ days.
Additionally, lags measured using the full six-year light curves exhibit a higher success rate compared to those derived from the seasonal one-year light curves. 
This is mainly because the long duration of the full six-year light curves minimizes the chances of random fluctuations in AGN variability creating spurious patterns that resemble observations, that is, the probability of $ r_{\rm max}^{\rm sim} > r_{\rm max}^{\rm obs} $ is significantly reduced.

\section{Discussion}\label{sec:results}

\subsection{A large fraction of AGN have variable lags}\label{sec:var_lag}

\begin{figure*}
\centering
\includegraphics[width=.32\linewidth]{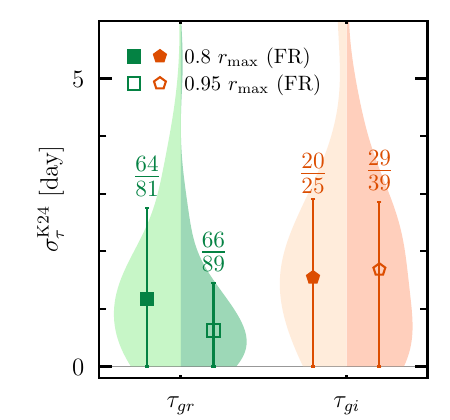}
\includegraphics[width=.32\linewidth]{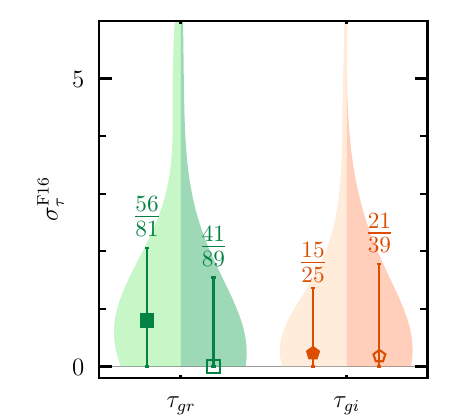}
\includegraphics[width=.32\linewidth]{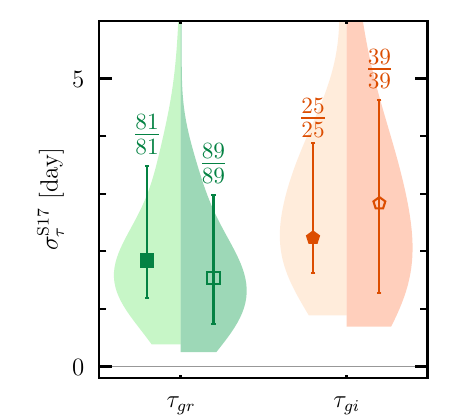}
\includegraphics[width=.32\linewidth]{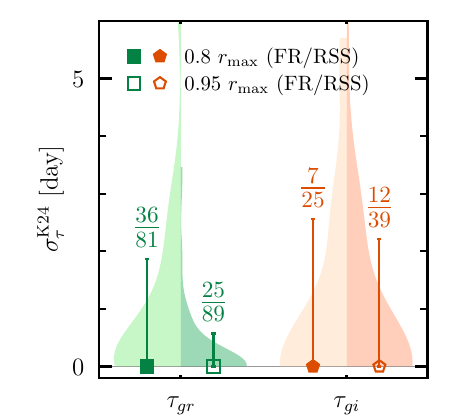}
\includegraphics[width=.32\linewidth]{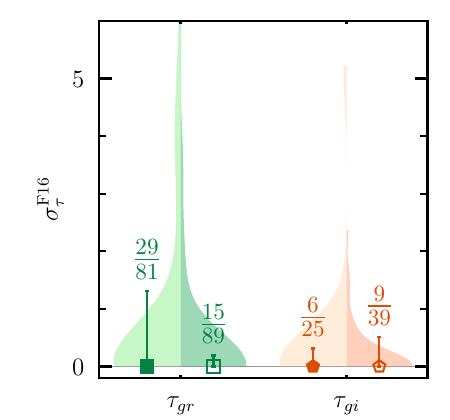}
\includegraphics[width=.32\linewidth]{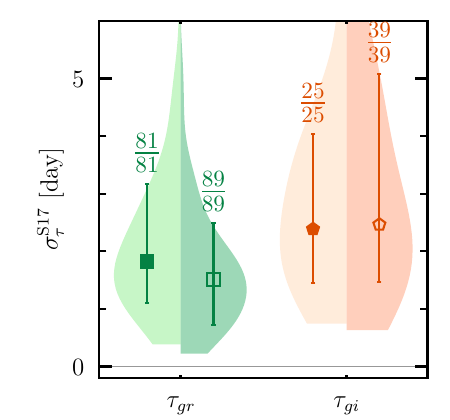}
\caption{
Top panels: distributions of the variation amplitude of lags, $\sigma_{\tau}$, for sub-samples with at least three successful lags of $\tau_{gr}$ or $\tau_{gi}$ across six seasons. Three distinct estimators, i.e., $\sigma_{\tau}^{\rm K24}$ (Equation~\ref{eq:var1}, left panel), $\sigma_{\tau}^{\rm F16}$ (Equation~\ref{eq:var2}, middle panel), and $\sigma_{\tau}^{\rm S17}$ (Equation~\ref{eq:var3}, right panel), are considered using the 1$\sigma$ FR uncertainties of lags.
In each panel, the left and right side of each violin plot represents the distribution of $\sigma_{\tau}$ for lags weighted with $r_{\rm cc} \geqslant 0.8~r_{\rm max}$ and $r_{\rm cc} \geqslant 0.95~r_{\rm max}$, respectively. Symbols (squares for $\tau_{gr}$ and pentagons for $\tau_{gi}$) are the median values overlaid by the 16\% to 84\% percentile range of the distribution.
Numbers of the sub-sample (denominator) and of those with $\sigma_{\tau} > 0$ (numerator) are marked.
It is clear that more than $50\%$ of AGN exhibit variable lags.
Bottom panels: same as the top panels, but $\sigma_{\tau}$ is estimated using the 1$\sigma$ FR/RSS uncertainties of lags.
Since the FR/RSS method tends to overestimate the lag uncertainties \citep{Yu2020}, the variation in lags reduces significantly as indicated by either $\sigma^{\rm K24}_{\tau}$ or $\sigma^{\rm F16}_{\tau}$ once over-subtracting the contribution of the lag uncertainties to the lag variation (see Section~\ref{sec:var_lag}).
}
\label{fig:agn_var_all}
\end{figure*}

Figure~\ref{fig:agn_lag} shows an individual AGN with variable lags across six seasons.
Now, we explore the statistical properties of the lag variation for the 94 AGN and mainly focus on the lag variations of $\tau_{gr}$ and $\tau_{gi}$ because $\tau_{gg}$ is always nearly identical to zero. Since every AGN has at most six successful lags measured in seasons, we require that each AGN should have at least three successful lags in order to quantify its lag variation.
For the centroid lags weighted with $r_{\rm cc} \geqslant 0.8~r_{\rm max}$, there are 81 and 25 AGN with at least three successful lags of $\tau_{gr}$ and $\tau_{gi}$, respectively. 
For the centroid lags weighted with $r_{\rm cc} \geqslant 0.95~r_{\rm max}$, numbers of AGN with at least three successful lags of $\tau_{gr}$ and $\tau_{gi}$ increase slightly to 89 and 39, respectively.
In this subsection, we would discuss the variation of the seasonal lags for each of these sub-samples.

For each AGN, we use the weighted excess variance, $\sigma^{\rm K24}_{\tau}$ (Equation~(\ref{eq:var1})), to quantify its variation in the seasonal lags. Since part of the observed variation in lags is attributed to the uncertainties in lags measured, the $\sigma^{\rm K24}_{\tau}$ estimator takes this into account by weighting and subtracting the contribution of the lag uncertainties to the lag variation. Therefore, the $\sigma^{\rm K24}_{\tau}$ estimator is expected to be sensitive to the lag uncertainties.
The top-left and bottom-left panels of Figure~\ref{fig:agn_var_all} present distributions of $\sigma^{\rm K24}_{\tau}$ for each sub-sample whose 1$\sigma$ lag uncertainties are estimated using the FR and FR/RSS methods, respectively.

In top-left panel of Figure~\ref{fig:agn_var_all}, when considering the FR uncertainties, $\sim 80\%$ of AGN with the centroid lags weighted with $r_{\rm cc} \geqslant 0.8~r_{\rm max}$ show variable lags (left side of the violin plot). A similar conclusion is obtained for $\sim 75\%$ of AGN with the centroid lags weighted with $r_{\rm cc} \geqslant 0.95~r_{\rm max}$ (right side of the violin plot).
Instead, in the bottom-left panel of Figure~\ref{fig:agn_var_all}, considering the FR/RSS uncertainties, only $\sim 30\%-40\%$ of AGN exhibit variable lags. However, the relatively lower fraction could be the natural consequence of over-subtracting the lag uncertainties since the FR/RSS method tends to overestimate the lag uncertainties \citep[e.g.,][]{Yu2020}.

Since properly weighting and subtracting the contribution of the lag uncertainties to the lag variation is indeed non-trivial, we adopt another commonly-used estimator, that is, the fractional variability \citep{Fau2016},
\begin{equation}
    \sigma_{\tau}^{\rm F16} = \sqrt{\left[
    \frac{1}{N-1}\sum_j(X_j-\bar{X})^2-\frac{1}{N}\sum_j\sigma_j^2\right]/\bar{X}^2,
    }\label{eq:var2}
\end{equation}
where $X_j$ and $\sigma_j$ are the lag and $1\sigma$ uncertainty in the $j$-th season and $\bar{X}$ is the weighted mean of the seasonal lags. 
Similar to the $\sigma^{\rm K24}_{\tau}$ estimator, the $\sigma_{\tau}^{\rm F16}$ estimator also accounts for the contribution of the lag uncertainties but subtracts the contribution in a different way and does not include weights according the lag uncertainties.
Note that the absolute values of the $\sigma^{\rm K24}_{\tau}$ and $\sigma_{\tau}^{\rm F16}$ estimators can not be compared because the former has a units of day while the latter is dimensionless.

Distributions of the variation amplitude of lags indicated by the $\sigma_{\tau}^{\rm F16}$ estimator are presented in the middle panels of Figure~\ref{fig:agn_var_all}. Similar to conclusions suggested by the $\sigma^{\rm K24}_{\tau}$ estimator, the $\sigma_{\tau}^{\rm F16}$ estimator suggest that $\sim 50\%-70\%$ of AGN exhibit variable lags if considering the FR uncertainties, while only $\sim 25\%-35\%$ of AGN have variable lags if considering the FR/RSS uncertainties. Again, the relatively lower fraction in the latter case could be owing to an over-subtraction of the contribution of the lag uncertainties to the lag variation.

Since both the $\sigma^{\rm K24}_{\tau}$ and $\sigma_{\tau}^{\rm F16}$ estimators may be subject to the problem of over-subtraction of the contribution of the lag uncertainties to the lag variation, especially for AGN with intrinsic minor variation in lags, we use one more estimator, that is, the weighted standard deviation \citep{Sokolovsky2017},
\begin{equation}
    \sigma_{\tau}^{\rm S17} = \sqrt{
    \frac{\sum \sigma_j^w}{\left( \sum \sigma_j^w \right)^2-\sum (\sigma_j^{2w})}\sum_j \sigma_j^w(X_j- \bar{X})^2
    } 
    \label{eq:var3}
\end{equation}
with $w=-2$. The $\sigma_{\tau}^{\rm S17}$ estimator does not explicitly excludes the contribution of the lag uncertainties to the lag variation but decrease to some extent that contribution by weights according to the lag uncertainties.
As illustrated in the right panels of Figure~\ref{fig:agn_var_all}, the $\sigma_{\tau}^{\rm S17}$ estimator suggest that all AGN have variable lags regardless of the ways of measuring the lags and estimating the lag uncertainties.

According to the above three estimators, we conservatively suggest that $\gtrsim50\%$ AGN exhibit variable lags.
It should be noted that none of the above estimators are robust enough. Although developing a robust estimator still deserves, a firm conclusion on the variation properties in lags of AGN would rely on future multi-wavelength high-precision high-cadence and long-term time domain surveys, such as those being conducted by WFST.

\subsection{What does the variable lag imply?}

\begin{figure*}[tbp]
    \centering
    \includegraphics[width=0.45\textwidth]{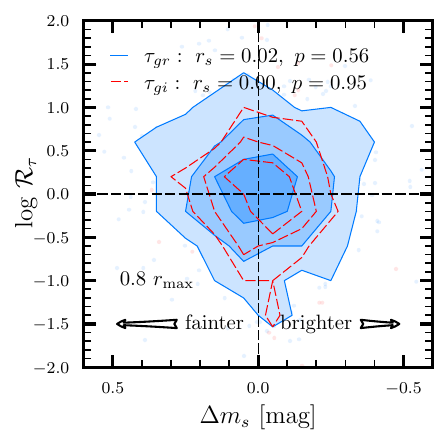}
    \includegraphics[width=0.45\textwidth]{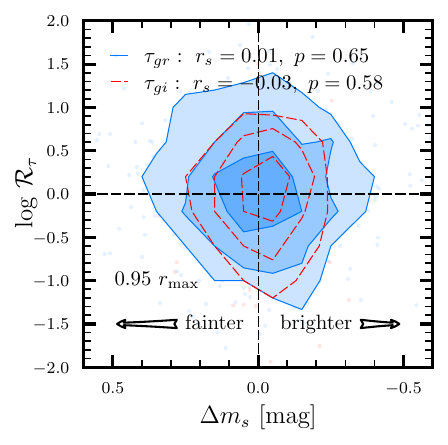}
    \caption{Left panel: contour plots of changes in the median magnitudes between two seasons ($\Delta m_s$) versus changes in the seasonal lags ($\mathcal{R_\tau}$) for $\tau_{gr}$ (blue solid contours) and $\tau_{gi}$ (red dashed contours) of the 94 AGN. The lags are derived using $r_{\rm cc} \geqslant 0.8~r_{\rm max}$. For both lags, the Spearman rank correlation coefficient ($r_{\rm s}$) and the associated significance level ($p$) indicate no significant correlation between $\Delta \tau_s$ and $\Delta m_s$.
    Right panel: same as the left one, but for lags derived using $r_{\rm cc} \geqslant 0.95~r_{\rm max}$.
    }
    \label{fig:dlag_dm}
\end{figure*}

If the lags indeed vary, there are several plausible interpretations: (1) the structural changes within the disk \citep{Yu2020}, (2) the dynamical evolution of the driving source, e.g., the evolution of the corona height, \citep{Kammoun2021}, (3) the varying contribution from clouds in BLR \citep{Vincentelli2021}, (4) the changes in the luminosity of AGN, e.g., larger interband lags when AGN get brighter \citep[][]{Sergeev2005,Zhou2024arXiv}, or (5) the inherent randomness of AGN variability \citep{Su2024b}.
While scenarios (1) to (3) could explain changes in lags, they require additional observational data for further confirmations. 
Thus, we mainly discuss whether changes in AGN luminosity (Section~\ref{sec:lag-luminosity}) and the inherent randomness of AGN variability (Section~\ref{sec:lag-random}) could be responsible for the observed variations in lags.

\subsubsection{Do lags change with luminosity?}\label{sec:lag-luminosity}

Recently, \cite{Zhou2024arXiv} reported a significant lag difference in the Seyfert galaxy NGC 4151 between two multiwavelength campaigns. Specifically, the lags measured in the high-flux state of NGC 4151 were about $3.8$ times larger than those in the low-flux state as its optical flux density increased from the low-flux state ($f_{\rm low}$) to the high-flux state ($f_{\rm high}$) by a factor of $\simeq3.3$.

To explore whether a similar trend presents in our sample of 94 AGN, we analyze the seasonal variations in both lag and median magnitude of light curves for each AGN.
We define the lag ratio as $\mathcal{R}_\tau=\tau_i/\tau_j$ and the median magnitude difference of light curves as $\Delta m_s=\mathrm{med}({m_{i}) - \mathrm{med}({m_{j}})}\equiv -2.5{\rm log}(f_i/f_j)$, where the subscripts $i$ and $j$ represent different observing seasons.

Figure~\ref{fig:dlag_dm} presents the contour plots of $\Delta m_s$ versus $\mathcal{R}_\tau$ for $\tau_{gr}$ and $\tau_{gi}$ of the 94 AGN.
Based on the current ZTF data, we find no significant correlation between lag variations and magnitude changes in these AGN.
The Spearman rank correlation coefficient is 0.02 (0.00) with a null hypothesis value of 0.56 (0.95) for $\tau_{gr}$ ($\tau_{gi}$) when using $r_{\rm cc} \geqslant 0.8~r_{\rm max}$. 
Similar results persist for $\tau_{gr}$ ($\tau_{gi}$) when using $r_{\rm cc} \geqslant 0.95~r_{\rm max}$, where the Spearman rank correlation coefficient is 0.01 (-0.03) with a null hypothesis value of 0.65 (0.58).

Acutally, a lag-luminosity relation, i.e., $\tau\propto L^{1/2}$, was first observed by \citet{Sergeev2005} using a sample of 14 AGN and later confirmed by \cite{GuoH2022} with a sample of 94 AGN.
The relation was also predicted by both the X-ray reprocessing model \citep[cf. Appendix~B in][]{GuoH2022} and the radiation-pressure confined BLR cloud model \citep{Netzer2022}.
In Figure~\ref{fig:dlag_dm}, the magnitude differences are measured for each individual AGN within the 94 AGN. Even considering the maximum magnitude difference, i.e., $\simeq0.7$ mag, would give rise to a variation in lags of $\simeq 0.14$ dex only, according to the lag-luminosity relation in an equivalent form of ${\rm log}(\mathcal{R}_\tau) \propto -0.2\Delta m_s$. Therefore, the lag variations we reported cannot be explained by changing the AGN luminosity.
Nonetheless, our results do not necessarily contradict the lag-luminosity relation because we focus on the magnitude difference of each individual AGN, rather than the magnitude difference of an AGN sample covering a broad range of magnitude, such as ranges of $\simeq 8$ mag in \citet{Sergeev2005} and $\simeq 10$ mag in \citet{GuoH2022}.

\subsubsection{Do lags change due to the inherent randomness of AGN variability?}\label{sec:lag-random}

\cite{Su2024b} recently considered two models for AGN variability for NGC 5548, i.e., the thermal fluctuation model and the reprocessing model with DCE from the BLR, and demonstrated that the inherent randomness of AGN variability can also naturally lead to changes in the interband lags inferred from a finite baseline, even without invoking any structural or external modifications.
In particular, for an individual AGN with repeated, nonoverlapping observations, the thermal fluctuation model generally predicts greater variations in lags compared to the reprocessing scenario. 
Moreover, in both scenarios, lags are expected to increase with lengthening the baseline and eventually saturate when the baseline of light curve becomes sufficiently long.

\begin{figure}[ht]
    \centering
    \includegraphics[width=0.22\textwidth]{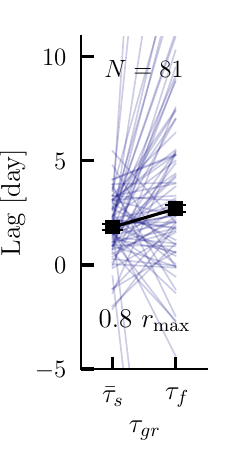}
    \includegraphics[width=0.22\textwidth]{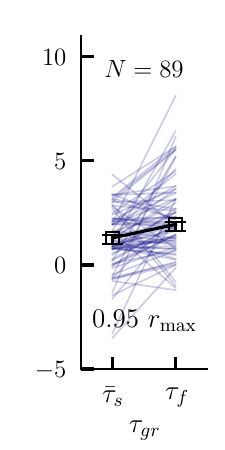}
    \includegraphics[width=0.22\textwidth]{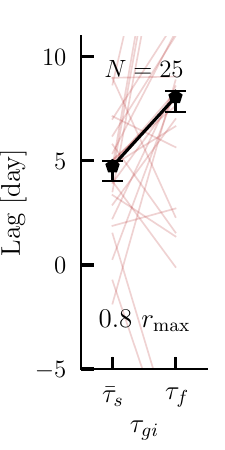}
    \includegraphics[width=0.22\textwidth]{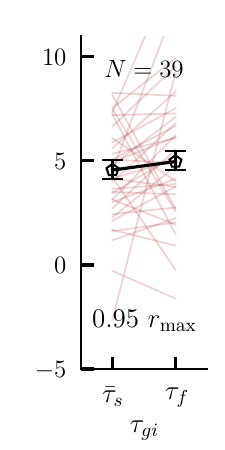}
    \caption{Comparison between the short-term lag, $\bar{\tau}_s$, inferred by averaging the seasonal one-year lags, and the long-term lag, $\tau_f$, inferred from the full six-year light curves, for those AGN with at least three successful seasonal lags (see the nominated $N$ for the corresponding sample size).
    The top and bottom panels are for $\tau_{gr}$ and $\tau_{gi}$, respectively. The left and right panels are for lags weighted with $r_{\rm cc} \geqslant 0.8~r_{\rm max}$ and $r_{\rm cc} \geqslant 0.95~r_{\rm max}$, respectively.
    Each thin line indicates the change from $\bar{\tau}_s$ to $\tau_f$ for each AGN, while symbols linked by a thick line are the resultant median values overlaid by the bootstrapped $1\sigma$ uncertainties.}
    \label{fig:lag_season_full}
\end{figure}

In Figure~\ref{fig:lag_season_full}, we display the dependence of the interband lag on the baseline of light curve using AGN with at least three successful lags across six seasons.
Adopting the centroid lags weighted with $r_{\rm cc} \geqslant 0.8~r_{\rm max}$, the short-term lags, $\bar{\tau}_s$, derived by averaging the seasonal lags, are generally smaller than the long-term lags, $\tau_f$, inferred from the full six-year light curves. 
This conclusion remains when adopting the centroid lags weighted with $r_{\rm cc} \geqslant 0.95~r_{\rm max}$.
Note that when adopting the centroid lags weighted with $r_{\rm cc} \geqslant 0.8~r_{\rm max}$, some AGN show anomalously large $\tau_{f}$, that is, up to $\simeq$50 days for $\tau_{gr}$ and up to $\simeq$20 days for $\tau_{gi}$. 
Visual inspection reveals broad and asymmetric CCFs for those AGN (see Appendix~\ref{appendix:ccf_shape} for two representative AGN), indicating that the lag measurements are sensitive to the choice of the weighting threshold, e.g., $\geqslant 0.8~r_{\rm max}$ or $\geqslant 0.95~r_{\rm max}$. 
Nonetheless, such anomalies are rare in the AGN sample.
In short, Figure~\ref{fig:lag_season_full} provides an evidence supporting that the lags measured are sensitive to the baseline of the light curve \citep[e.g.,][]{Su2024b}. Note that for the two lag measurements in NGC 4151 reported by \cite{Zhou2024arXiv}, the larger lag also corresponds to a longer baseline, consistent with the pattern observed in Figure~\ref{fig:lag_season_full}.

Both the significant changes in the lag ratio with small magnitude variations (Figure~\ref{fig:dlag_dm}) and the tendency of increasing lags with the baseline of light curve (Figure~\ref{fig:lag_season_full}) suggest that the lag variations may simply arise from the inherent randomness of AGN variability.
Further quantitative comparisons between two models (i.e., the thermal ﬂuctuation model and the reprocessing model with DCE from the BLR) and the ZTF data are being carried out for AGN with different BH masses and luminosities (Z. B. Su et al. 2025, in preparation).

\subsubsection{Towards a better understanding of lags measured}

To understand the underlying physical mechanisms of the interband lag as well as its variation, robustly measuring the lag is essential. 
Over the years, many techniques have been developed to estimate the lag, including (1) the traditional CCF method \citep{Peterson2005, PyCCF2018}, (2) \texttt{JAVELIN} \citep{Zu2011}, and (3) \texttt{PYROA} \citep{Donnan2021}.
However, the use of inappropriate or over-simplified methods could lead to misinterpretations. 
For instance, both the traditional CCF method and \texttt{JAVELIN} may produce unreliable results if the assumption of a linear CRM model does not hold \citep{Yu2020b}. 
Moreover, \texttt{JAVELIN} models the light curve by assuming the DRW process. This assumption has been shown to be inconsistent with observations on both very short timescales ($\sim$minutes; \citealt{Mushotzky2011,Su2024a}) and very long timescales ($\sim$years to decades; \citealt{Guo2017}).
In contrast, the \texttt{PYROA} method combines all observed light curves to build a running optimal average (ROA) model that can be normalized, shifted, and scaled to fit each individual observed light curve. 
The \texttt{PYROA} method allows for simultaneously measuring the interband lag as well as both the mean and rms variability.  
However, the validation of the \texttt{PYROA} method remains unclear when the observed light curves is not the simply shifted version of the ROA model, especially when the light curves exists some anomalous features \citep[e.g.,][]{Gaskell2021}.
Other more advanced methods, such as the frequency-resolved lags \citep[e.g.,][]{Cackett2022ApJ} and the wavelet transform method \citep{Wilkins2023}, may also help unveil deeper properties of the interband lags when multi-wavelength quasi-simultaneous high-cadence observations are available.

The variable lag reported here raises important questions on our understanding of the interband lags. 
Specifically, the current CRM projects, which typically span short (e.g., $\lesssim$ 1 year) durations, lack repeated monitoring, and often do not provide sufficiently long light curves\footnote{The required monitoring length might also depend on AGN properties \citep[e.g.,][]{Chen2024}.}, may not fully capture the true nature of the interband lags such as its variation properties.
As a result, relationships between lags and physical properties of AGN may require further revision.
In the near future, the high-precision photometry surveys would clarify whether to what extent there is a scatter in the lag as well as its relationship to physical properties of AGN. 
Thereafter, a model for AGN variability should be able to simultaneously address these properties of the interband lag.

\begin{figure*}[ht]
    \centering
    \includegraphics[width=0.45\textwidth]{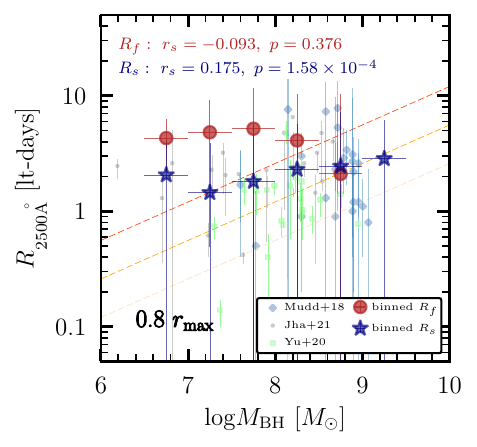}
    \includegraphics[width=0.45\textwidth]{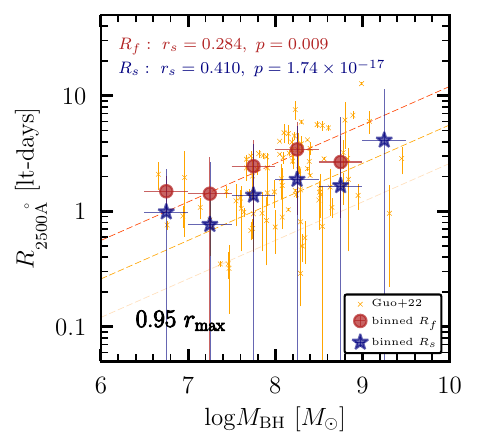}
    \caption{
    Correlations between 2500~\angstrom\ disk size and BH mass. The disk size is inferred from the centroid lags weighted with $r_{\rm cc} \geqslant 0.8~r_{\rm max}$ (left panel) or $r_{\rm cc} \geqslant 0.95~r_{\rm max}$ (right panel).
    Using the seasonal one-year and full six-year light curves, asterisks ($R_s$) and circles ($R_f$) indicate the average disk sizes in BH mass bins of 0.5 dex, respectively. 
    The Spearman rank correlation coefficient ($r_s$) and the significance level ($p$) are nominated for correlations between disk size and BH mass.
    In the left panel, we also present the disk size measurements by \citet[][diamonds]{Mudd2018}, \citet[][squares]{Yu2020}, and \citet[][dots]{Jha2022}, while in the right panel those by \citet[][crosses]{GuoH2022} using the same $r_{\rm cc} \geqslant 0.95~r_{\rm max}$.
    In both panels, three dashed lines from top to bottom represent the prediction of the SSD model for Eddington ratios of 1, 0.1, and 0.01, respectively. Note that for each AGN, we have $\leqslant 6$ measurements of disk size $R_s$ but only one measurement of $R_f$, resulting in a significantly smaller null hypothesis value for $R_s$ compared to $R_f$.}
    \label{fig:lag_mbh}
\end{figure*}

\subsection{The relationship between disk size and BH mass}\label{sec:lag_mbh}

The SSD model predicts that the characteristic radius, $R_{\rm SSD, \lambda}$, for emission at rest-frame wavelength $\lambda$ scales with BH mass and Eddington ratio as \( R_{\rm SSD, \lambda} \propto M_{\rm BH}^{2/3} \lambda_{\rm Edd}^{1/3} \lambda^{4/3} \) \citep[e.g.,][]{MacLeod2010}. 
This relation implies a positive correlation between disk size and BH mass. However, several observational results have so far yielded inconsistent conclusions.

Studies which used the {thin-disk model} implemented in \texttt{JAVELIN} \citep{Mudd2018} to estimate the disk size consistently reported no significant correlation between disk size and BH mass \citep[e.g.,][]{Mudd2018,Yu2020,Jha2022}.
In contrast, \cite{GuoH2022} estimated the disk size from the interband lags weighted with $r_{\rm cc} \geqslant 0.95~r_{\rm max}$ and reported a strong correlation between disk size and BH mass. 
More recently, \cite{Sharp2024} measured the \texttt{JAVELIN} lags for a sample of 95 luminous quasars from the SDSS-RM project and found an anti-correlation between the ratio of the observed $\texttt{JAVELIN}$ lag to the SSD-predicted lag and BH mass.
These controversial findings highlight the complexity and uncertainty in the observed relation between disk size and BH mass. 
Therefore, we revisit the dependence using the 94 AGN and explore whether differences in the threshold used to measure the lags might influence the observed relation between disk size and BH mass.

For each AGN, the 2500 \angstrom\ disk size is simply estimated using the SSD-predicted scaling relation, $R_{2500\mathring{\rm A}} = R_g \left( \lambda_{2500\mathring{\rm A}} / \lambda_g \right)^{4/3}$, where the $g$-band disk size $R_g = c\tau_g$ and $\tau_g$ comes from the best-fit lag-wavelength relation (see Section~\ref{sec:amplitude_lag}). 
Since $\tau_g$ can be measured in two ways, i.e., either using seasonal one-year light curves or using full six-year light curves, we would have two kind of disk sizes, $R_s$ or $R_f$, respectively.

Figure~\ref{fig:lag_mbh} displays the BH-mass-bin-averaged 2500 \angstrom\ disk size versus the BH mass for the 94 AGN. The left and right panels of Figure~\ref{fig:lag_mbh} present the disk sizes  inferred from the centroid lags weighted with $r_{\rm cc}\geqslant 0.8~r_{\rm max}$ and $r_{\rm cc}\geqslant 0.95~r_{\rm max}$, respectively.
Note that each AGN have $\leqslant 6$ disk sizes, $R_s$, but only one $R_f$.
In both panels, $R_f$ tends to be systematically larger than $R_s$, again indicating that the length of light curve has an influence on the lag measurement (see Section~\ref{sec:lag-random}).
For comparison, assuming $\eta=0.1$ and $X=5.04$ \citep{Tie2018}, the SSD-predicted disk sizes are overlaid for various Eddington ratios.

Illustrated in the left panel of Figure~\ref{fig:lag_mbh} where disk sizes are inferred from lags weighted with $r_{\rm cc}\geqslant0.8~r_{\rm max}$, a weak if any correlation is found between disk size and BH mass. The Spearman rank correlation coefficients are 0.175 for $R_s$ and -0.093 for $R_f$, with corresponding null hypothesis values of $1.5\times10^{-4}$ and 0.376, respectively. 
In contrast, the right panel of Figure~\ref{fig:lag_mbh}, where the disk sizes are inferred from lags weighted with $r_{\rm cc}\geqslant0.95~r_{\rm max}$, reveals a stronger correlation between disk size and BH mass.
In this case, the Spearman rank correlation coefficients increase to 0.410 for $R_s$ and 0.284 for $R_f$, with much lower null hypothesis values of $1.74\times10^{-17}$ and 0.009, respectively.
Our results help reconcile the discrepancy between the findings reported by \cite{GuoH2022} and others \citep[e.g.,][]{Yu2020,Jha2022}. 
While \cite{GuoH2022} attributed the discrepancy to several factors, e.g., stricter criteria in sample selection, lag selection, or the high-quality light curves used, our results suggest that the major reason should be the usage of distinct thresholds to measure the centroid lags.

The disk sizes inferred from the interband lags are known to conflict with, typically larger than, predictions from the SSD model. 
Both Figures~\ref{fig:lag_season_full} and~\ref{fig:lag_mbh} demonstrate that the lags measured depend on the baseline of light curve, consistent with recent simulations \citep{Chen2024,Su2024b}. 
This suggests that the choice of the baseline can influence the lags measured and the associated interpretations. 
For example, \cite{Wang2023} found that the size of BLR is correlated with, and on average $\simeq$8.1 times larger than, the size of the continuum-emitting region at 5100 \angstrom. 
They argued that this correlation is an alternative for estimating BH mass.
% , particularly in the high-cadence photometric era.
However, if the interband lags tend to increase with the length of the light curve, systematic uncertainties on the BH mass are unavoidable when simply applying such a correlation.

In short, we suggest that both the length of light curve and the way of measuring the lag should be taken carefully before concluding whether there is a correlation between disk size and BH mass. 
Alternative approaches, such as microlensing \citep[e.g.,][]{Bate2018MNRAS.479.4796B,Morgan2018}, could provide independent constraints on disk size and thus offer a complementary test on this correlation.

\section{Conclusions}\label{sec:cons}

In this work, using both the seasonal one-year and full six-year $gri$-band light curves from the ZTF DR22, we analyze the interband lags of a sample of 94 AGN with significant lag measurements suggested by \cite{GuoH2022}.
Our findings are as follows:

\begin{itemize}

\item[1.] For many individual AGN, the interband lags show clear variation across six observing seasons, which may simply arise from the inherent randomness of AGN variability.

\item[2.] The lag measurement is sensitive to the baseline of the light curve. Short-term lags, derived by averaging lags inferred from multiple seasonal light curves, are typically smaller than the long-term lags, which are inferred from the full six-year light curves. This behavior is consistent with recent simulations.

\item[3.] The disk size, inferred from the centroid lags weighted with $r_{\rm cc} \geqslant 0.8~r_{\rm max}$, only has a weak if any correlation with the BH mass. Instead, using the centroid lags weighted with $r_{\rm cc} \geqslant 0.95~r_{\rm max}$, a stronger correlation between the inferred disk size and the BH mass is observed.

\end{itemize}

Our findings suggest that the lags are more variable and complex than previously thought.
A better understanding of the origin of interband lags requires accurate and repeated lag measurements.
In the era of high-precision time-domain surveys, such as those by WFST and LSST, both deep detection limits and high-precision photometry will offer such an opportunity to shed new light on the physical origin of the AGN lag.

\section*{Acknowledgement}

This work is supported by the National Key R\&D Program of China (grant No. 2023YFA1608100) and the National Natural Science Foundation of China (grant Nos. 12373016, 12473018, and 12033006).
M.Y.S. acknowledges support from the National Natural Science Foundation of China (NSFC-12322303), and the Natural Science Foundation of Fujian Province of China (No. 2022J06002). 

Based on observations obtained with the Samuel Oschin 48 inch Telescope at the Palomar Observatory as part of the Zwicky Transient Facility project. Z.T.F. is supported by the National Science Foundation under grant No. AST-1440341 and a collaboration including Caltech, IPAC, the Weizmann Institute for Science, the Oskar Klein Center at Stockholm University, the University of Maryland, the University of Washington, Deutsches Elektronen-Synchrotron and Humboldt University, Los Alamos National Laboratories, the TANGO Consortium of Taiwan, the University of Wisconsin at Milwaukee, and Lawrence Berkeley National Laboratories. Operations are conducted by COO, IPAC, and UW.

\facility{ZTF\citep{IRSA_ZTF}.}

\appendix
\section{AGN with Broad, Asymmetric CCF}\label{appendix:ccf_shape}
\begin{figure*}
    \centering
    \includegraphics[width=0.95\textwidth]{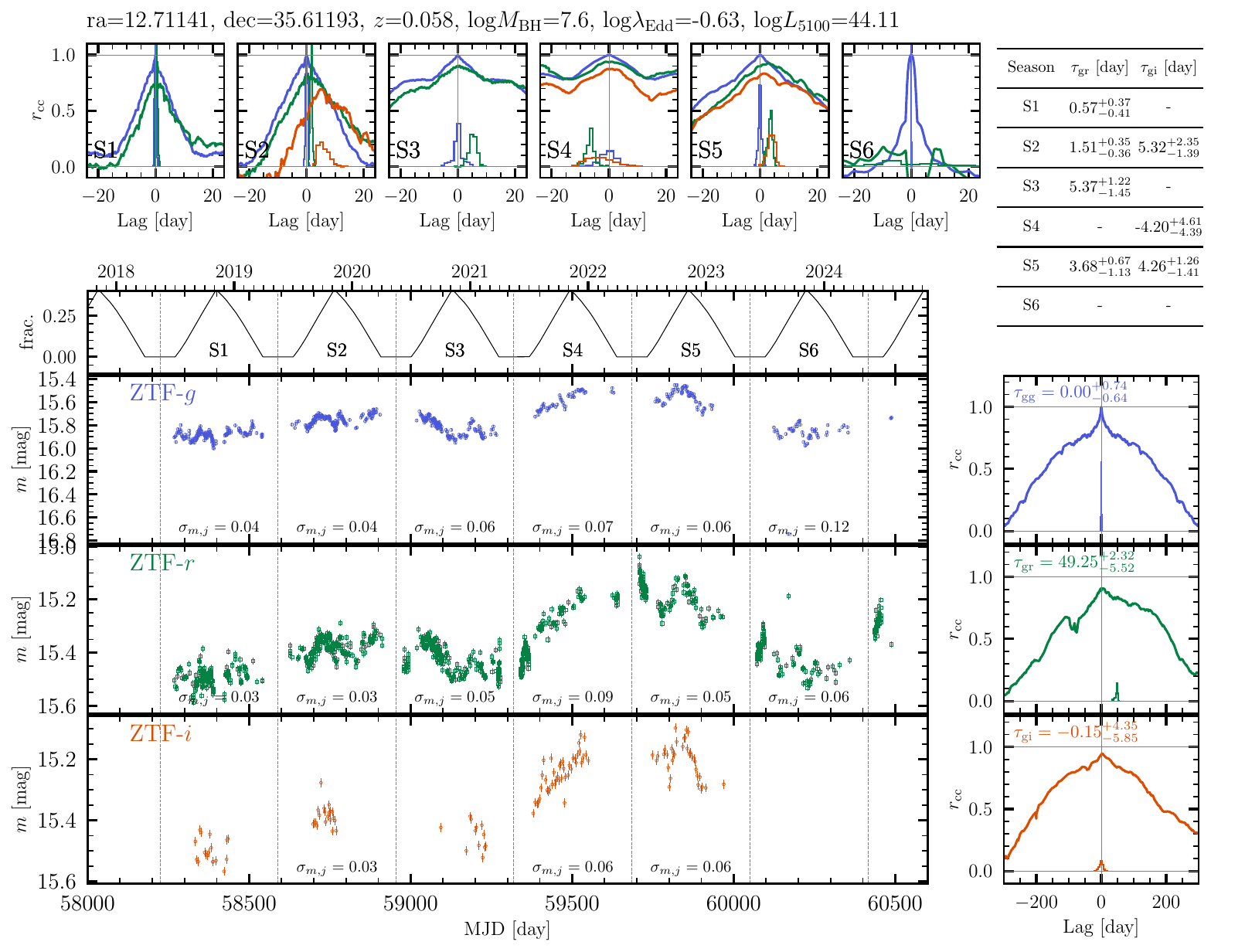}
    \caption{Same as Figure~\ref{fig:ztf_lc_ccf}, but for an AGN possessing a very large long-term lag $\tau_{gr}$, as a result of the notably broad and asymmetric CCF. Note that the $x$-axis ranges of the right-column panels have been expanded to $\pm300$ days to display the full shape of the CCF.}\label{fig:appendix1}
\end{figure*}

\begin{figure*}
    \centering
    \includegraphics[width=0.95\textwidth]{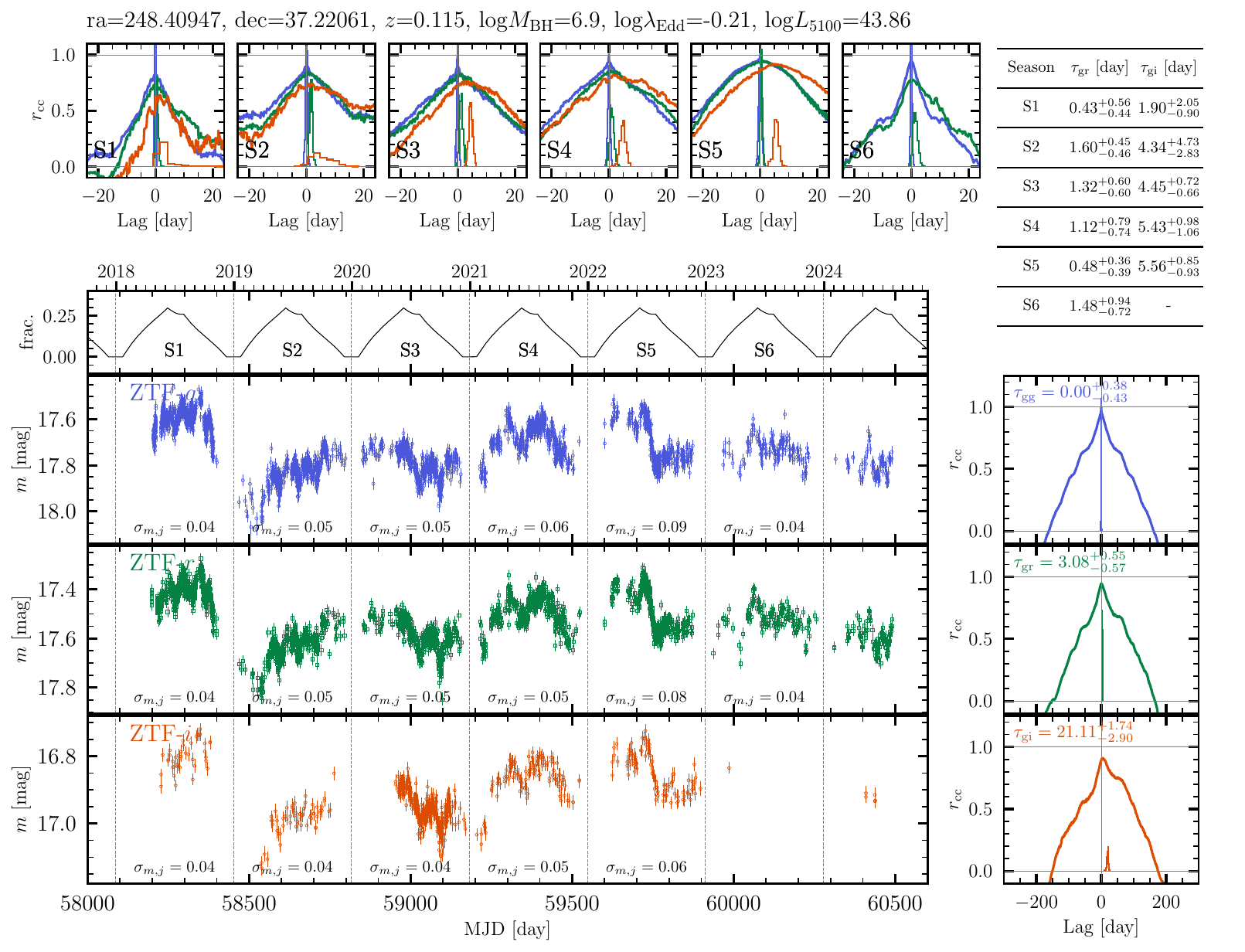}
    \caption{Same as Figure~\ref{fig:appendix1}, but for an AGN possessing a very large long-term lag $\tau_{gi}$.
    }\label{fig:appendix2}
\end{figure*}

As shown in Figure~\ref{fig:lag_season_full}, we find that some AGN have unexpectedly large long-term lags ($\tau_f$) inferred from the full six-year light curves. 
In Figures~\ref{fig:appendix1} and \ref{fig:appendix2}, we illustrate the ZTF $gri$-band light curves and the corresponding CCF results for two AGN with very large long-term lags of $\tau_{gr} \sim 50$ days and $\tau_{gi} \sim 20$ days, respectively. 
For both cases the maximum correlation coefficients, $r_{\rm max}$, are $\simeq 0.91$ with significantly low values of $p(r_{\rm max})$. 
Their broad and asymmetric CCFs results in the very large long-term lags when adopting a threshold of $\geqslant 0.8~r_{\rm max}$. If adopting a higher threshold of $\geqslant 0.95~r_{\rm max}$ which is less sensitive to the asymmetry of the CCF, much smaller long-term lags would be obtained. For example, the long-term $\tau_{gr}$ decreases from $49.25^{+5.84}_{-2.46}$ (Figures~\ref{fig:appendix1}) to $6.16^{+0.40}_{-0.30}$ days, while the long-term $\tau_{gi}$ decreases from $21.11^{+3.24}_{-1.94}$ (Figures~\ref{fig:appendix2}) to $5.54^{+0.41}_{-0.50}$ days.
These results highlight the uncertainties induced by the threshold in lag estimation when CCF is not asymmetric. 
Developing a robust method for the lag estimation is warranted but is beyond the scope of this paper.
Nonetheless, such very large long-term lags are rare among the 94 AGN, and the conclusion drawn from Figure~\ref{fig:lag_season_full} remains.


\bibliographystyle{aasjournal}
\bibliography{sample}

\end{document}